\def\rn{\noindent\parshape 2 0truecm 8.5truecm 0.3truecm 8.2truecm}
\def\rn{}
\def\nn#1 #2{#2. #1}				
\def\nnn#1 #2 #3{#2. #3. #1}			
\def\nnnn#1 #2 #3 #4{#2. #3. #4 #1}		
\def\nnnnn#1 #2 #3 #4 #5{#2. #3. #4 #5. #1}	
\def\dualand{ and\hbox{ }}				
\def\multiand{, and\hbox{ }}				
\def\rf#1;#2;#3;#4;#5 {{\frenchspacing\par\rn#1, #3 {\bf #4}, #5 (#2). \par}}
\def\rg#1;#2;#3;#4;#5;#6 {{\frenchspacing\par\rn#1, #3 {\bf #4}, #5 (#2). \par}}
\def\rfbook#1;#2;#3;#4;#5 {{\frenchspacing\par\rn#1, {\it #3} (#5, #4, #2).\par}}
\def\rfproc#1;#2;#3;#4;#5;#6 {{\frenchspacing\par\rn#1 #2, in {\it #3}, ed. #4 (#5: #6)\par}}
\def\rfprep#1;#2;#3 {{\par\frenchspacing\rn#1, #3 (#2).\par}}

\def\preskip{\vskip-1.3cm}
\def\postskip{\vskip-0.5cm}

\def\expec#1{\langle#1\rangle}

\def\etal{{\frenchspacing\it et al.}}
\def\ie{{\frenchspacing\it i.e.}}
\def\eg{{\frenchspacing\it e.g.}}

\def\crr{\cr\noalign{\vskip 4pt}}

\def\beq#1{\begin{equation}\label{#1}}
\def\eeq{\end{equation}}
\def\beqa#1{\begin{eqnarray}\label{#1}}
\def\eeqa{\end{eqnarray}}
\def\eq#1{equation~(\ref{#1})}
\def\Eq#1{Equation~(\ref{#1})}
\def\eqn#1{~(\ref{#1})}

\def\fig#1{Figure~\ref{#1}}
\def\Fig#1{Figure~\ref{#1}}

\def\sec#1{Section~\ref{#1}}

\def\spose#1{\hbox to 0pt{#1\hss}}
\def\simlt{\mathrel{\spose{\lower 3pt\hbox{$\mathchar"218$}}
     \raise 2.0pt\hbox{$\mathchar"13C$}}}
\def\simgt{\mathrel{\spose{\lower 3pt\hbox{$\mathchar"218$}}
     \raise 2.0pt\hbox{$\mathchar"13E$}}}
\def\simpropto{\mathrel{\spose{\lower 3pt\hbox{$\mathchar"218$}}
     \raise 2.0pt\hbox{$\propto$}}}

\def\ed{\end{document}}

\def\draft{
}

\def\lmax{{\ell_{\rm max}}}
\def\Nb{{N_{b}}}
\def\norm{{\cal N}}
\def\dT{\delta T}
\def\tr{\hbox{tr}\>}

\def\p{{\bf p}}
\def\n{{\bf n}}
\def\q{{\bf q}}

\def\xt{\tilde{\bf x}}
\def\x{{\bf x}}
\def\y{{\bf y}}
\def\z{{\bf z}}
\def\rh{\widehat{\bf r}}

\def\zh{\widehat{\z}}
\def\zero{{\bf 0}}

\def\A{{\bf A}}
\def\B{{\bf B}}
\def\C{{\bf C}}

\def\I{{\bf I}}
\def\F{{\bf F}}

\def\M{{\bf M}}
\def\N{{\bf N}}
\def\NN{{\bf\Sigma}}
\def\L{{\bf L}}
\def\Lt{{\bf\tilde L}}

\def\P{{\bf P}}
\def\Q{{\bf Q}}
\def\U{{\bf U}}
\def\R{{\bf R}}

\def\T{{\bf T}}
\def\W{{\bf W}}

\def\fsky{f_{\rm sky}}
\def\l{\ell}

\def\dT{{\delta T^2}}\def\ith{i^{th}}

\documentstyle[prd,aps,epsf]{revtex}
\draft
\begin{document}
\twocolumn[\hsize\textwidth\columnwidth\hsize\csname@twocolumnfalse\endcsname


\title{How to measure CMB polarization power spectra without losing information}

\author{Max Tegmark \& Angelica de Oliveira-Costa}


\address{Dept. of Physics, Univ. of Pennsylvania, 
Philadelphia, PA 19104; max@physics.upenn.edu}

\date{Submitted to Phys. Rev. D. December 7 2000, accepted February 15 2001}

\maketitle

\begin{abstract}
We present a method for measuring CMB polarization power spectra 
given incomplete sky coverage and test it with simulated
examples such as Boomerang 2001 and MAP.
By augmenting the quadratic estimator method
with an additional step, we find that the $E$ and $B$ power spectra 
can be effectively disentangled on angular scales 
substantially smaller than the width of the sky patch in the 
narrowest direction.
We find that the basic quadratic and maximum-likelihood methods
display a unneccesary sensitivity to systematic errors when
$T-E$ cross-correlation is involved, and show how this problem can be
eliminated at negligible cost in increased error bars.
We also test numerically the widely used approximation
that sample variance scales inversely with sky coverage,
and find it to be an excellent approximation on scales
substantially smaller than the sky patch.
\end{abstract}


] 


\section{Introduction}

As experimental groups roar ahead to map the CMB intensity with
increasing resolution and sensitivity, a second parallel 
front is being opened up: CMB polarization.
Theoretical advances now allow 
model predictions for CMB polarization to be computed with exquisite
accuracy 
\cite{Kamion97,Zalda97,Hu97},
and it is known that 
polarization measurements can substantially improve
the accuracy with which parameters can be measured over the
no-polarization case by breaking the degeneracy between 
certain parameter combinations \cite{ZSS97,parameters2,foregpars}.
CMB polarization can also provide 
crucial cross-checks and tests of the underlying theory. 
After placing ever tighter upper limits \cite{Hedman,StaggsReview},
a number of experimental teams are likely to make the first 
detections of CMB polarization in the next year or two,
about a decade after unpolarized CMB fluctuations were first 
detected \cite{Smoot92}. It is therefore timely to address
real-world issues related to extracting polarized power spectra from
experiments with incomplete sky coverage and complicated noise.
This is the purpose of the present paper, with emphasis on the problem
of separating the two polarization signals known as $E$ and $B$
\cite{Kamion97,Zalda97}. Important steps in this direction were 
first taken in \cite{Keating98,Zalda98}, for the special case of 
experiments mapping circles in the sky.
 
The $E/B$ problem goes back to the fact that the polarization tensor field
on the sky can be separated into a curl-free and a divergence free component,
and is most naturally expressed in terms of two scalar fields,
denoted $E$ and $B$ by analogy with electromagnetism \cite{Kamion97,Zalda97}.
Not only does this separation eliminate the coordinate system 
dependence that plagues the familiar Stokes parameters,
but $E$ and $B$ also probe distinct physical effects, 
making them the natural meeting point for theory and observation.
Unfortunately, the correspondence between $E$ and $B$ and the measured 
Stokes parameters is not spatially local, involving a partial 
differential equation, which means that
it is not possible to uniquely recover $E$ and $B$ from a map with 
merely partial sky coverage. This issue is particularly important
since the $B$-signal from inflation-produced gravity waves
potentially offers a unique probe of ultra-high-energy physics, but
may be swamped by a leakage from a larger $E$-signal 
\cite{Peterson00,Kamion00}.

A key goal of CMB analysis is to constrain cosmological models, 
and information-theoretical methods have been frequently
employed in the literature to study how accurately this can 
be done in principle with a given data set, using the 
Fisher information matrix formalism \cite{Fisher,karhunen}.
For CMB polarization experiments, this has been useful both
for optimizing experimental design \cite{JaffePursuer,KosowskyPursuer}
and for accuracy forecasting in general \cite{ZSS97}.
These information-theoretical tools are equally useful for data analysis, 
since they provide a simple way of checking whether cosmological
information is being lost in the data analysis pipeline.
Each step in such a pipeline typically compresses
the input data into a smaller set of numbers, and if
the output can be shown to retain all the cosmological information
of the input, the method is said to be lossless.
Lossless methods have been developed and extensively tested for the
unpolarized case, for both mapmaking \cite{mapmaking,Wright96b}
and power spectrum estimation \cite{cl,BJK}. As we will see,
it is possible to draw heavily on these methods for the 
polarized case as well, although a number of adaptations make
them simpler to interpret and help
improve the $E/B$-separation and robustness
towards systematic errors.

The rest of this paper is organized as follows.
In \sec{MethodSec}, we discuss basic methods involved in
polarization data analysis. To keep the presentation from becoming 
too abstract, many method details and extensions are 
postponed to \sec{ResultsSec}, where they are illustrated with
plots from actual applications. This section also assesses
the effectiveness of the various methods numerically, 
with applications to five experimental examples.
A step-by-step
summary of how to compute the signal covariance matrix is 
given in Appendix A for the reader wishing to do so in practice.
Our conclusions are summarized in \sec{DiscussionSec}.

\section{Method}
\label{MethodSec}

In this section, we first establish some basic notation, 
then discuss the extraction of maps from raw time-ordered data, 
and finally cover the extraction of power spectra from maps.

\subsection{Notation}

The linear polarization pattern of the CMB sky is 
characterized by the two Stokes parameters $Q$ and $U$ in each sky direction. 
$Q$ and $U$ are the components of a rank 2 tensor (spinor), 
loosely speaking a vector without an arrow on it,
so $Q$ and $U$ maps are defined only relative to a convention providing 
a reference direction at each point in the sky.
Let us discretize the $T$, 
$Q$ and $U$ maps into $N$ pixels centered at 
unit vectors
$\rh_1,...,\rh_N$, and write $T_i=T(\rh_i)$, $Q_i=Q(\rh_i)$, $U_i=U(\rh_i)$
(we let $T$ denote the unpolarized intensity, often called $I$).
We group these numbers into three $N$-dimensional vectors $\T$, $\Q$ and $\U$
and group these in turn into a single $3N$-dimensional vector
\beq{xDefEq}
\x\equiv\left(\begin{tabular}{c}
$\T$\\
$\Q$\\
$\U$
\end{tabular}\right).
\eeq
The statistical properties of $\x$ have been computed in 
full detail in the literature \cite{Kamion97,Zalda97},
and are characterized by six separate power spectra:
$C_\l^T$ for the unpolarized signal $T$, $C_\l^E$ for the
$E$-polarization, 
$C_\l^B$ for the $B$-polarization, and
$C_\l^{TE}$, $C_\l^{TB}$ and $C_\l^{EB}$ for the three
possible cross-correlations.
The power spectra $C_\l^{TB}$ and $C_\l^{EB}$ are both predicted 
to vanish for the CMB, but it will be interesting to measure them 
nonetheless, as probes of polarized foregrounds 
and exotic parity-violating physics \cite{Lue99}.
We will occasionally refer to the three cross-correlations $(TE,TB,EB)$ as
$(X,Y,Z)$, respectively.
We will find it useful for data analysis purposes
to recast the polarization problem in exactly the same
mathematical form as the simpler unpolarized case,
encoding all complications in a set of matrices.
The vector $\x$ has a vanishing expectation value 
(${\expec{\x}=\zero}$), and we can write its covariance matrix
as
\beq{CdefEq}
\C\equiv \expec{\x\x^t}= \sum_i p_i \P_i
\eeq
for a set of parameters $p_i$ and known matrices $\P_i$.
If the six observed power spectra are negligibly 
small 
for all multipoles $\l$ above some value $\lmax$ (which is 
always the case because of the smoothing caused by the 
finite angular resolution of an experiment), 
then we define these parameters to be
\beq{pDefEq}
\cases{
p_1,...,p_{\lmax-1} 		&$=\dT^T_2,..\dT^T_\lmax$,\crr
p_{\lmax},...,p_{2\lmax-2} 	&$=\dT^E_2,..\dT^E_\lmax$,\crr
p_{2\lmax-1},...,p_{3\lmax-3}	&$=\dT^B_2,..\dT^B_\lmax$,\crr
p_{3\lmax-2},...,p_{4\lmax-4}	&$=\dT^{TE}_2,..\dT^{TE}_\lmax$,\crr
p_{4\lmax-3},...,p_{5\lmax-5}	&$=\dT^{TB}_2,..\dT^{TB}_\lmax$,\crr
p_{5\lmax-4},...,p_{6\lmax-6}	&$=\dT^{EB}_2,..\dT^{EB}_\lmax$,\crr
p_{6\lmax-5}			&$=\eta=1.$
}
\eeq
\goodbreak

\noindent
Here
\beq{dTdefEq}
\dT_\l^P\equiv {\l(\l+1)\over 2\pi}C^P_\l
\eeq
is the familiar rescaled power that is normally used in power spectrum 
plots in place of $C^P_\l$. The index $P$ denotes any of the six
power spectrum types, \ie, $P=T,E,B,TE,TB,EB$.
The parameter $\eta$ is the normalization of the detector noise in the maps
relative to the predicted value, and is normally equal to unity.
Since $\C$ depends linearly on the parameters $p_i$, we can define
the $\P$-matrices as
\beq{PdefEq}
\P_i\equiv {\partial\C\over\partial p_i}.
\eeq
In other words, the first $6(\lmax-1)$ $\P$-matrices
give the contributions from the 
$T$, $E$, $B$, $TE$, $TB$ and $EB$ power spectra, and the last one is
simply the noise covariance matrix, giving the contribution from 
experimental noise.

Appendix A summarizes how to compute the $\C$-matrix,
and is intended for the reader who wishes to 
write software to do this in practice.
The $\P$-matrix corresponding to 
$\dT_\l^P$ is obtained from these formulas by simply setting 
all power spectra to zero, with the single exception
$\dT_\l^P=1$, \ie, $C^P_\l = 2\pi/\l(\l+1)$.
These $\P$-matrices are therefore independent of the
actual power spectra, and depend merely on the relative orientations
of the map pixels.

\subsection{Background: from timestream to $T$, $Q$ \& $U$ maps}
\label{mapSec}

To place our problem in context, this section briefly 
reviews how to reduce experimental
data to maps in the form of \eq{xDefEq}.
Although it is widely known how to do this, 
we will see that there are some subtle 
issues related to unmeasured modes.

\subsubsection{The basic inversion}

Suppose we have observed the sky a large number of times
with (perhaps) polarized detectors in a variety of
different orientations.
Let $y_i$ denote the number measured in the $\ith$ observation,
and group this time-ordered data set (TOD) into an $M$-dimensional 
vector $\y$.
The observed temperature fluctuation $y_i$ 
seen through a linear polarizer takes the form 
\beq{sincosEq1}
y_i = {1\over 2}\left[T_{k_i} + Q_{k_i} \cos(2\alpha_i) + U_{k_i} \sin(2\alpha_i)\right] + n_i,
\eeq
where $\alpha_i$ gives the clockwise angle between the polarizer and the reference
direction of the coordinate system, $n_i$ denotes the detector noise,
and $k_i$ is the number of the pixel pointed to during the $\ith$ observation.
If the experiment instead measures the difference between two perpendicular
polarizations, $y_i$ takes the simpler form 
\beq{sincosEq2}
y_i = Q_{k_i} \cos(2\alpha_i) + U_{k_i} \sin(2\alpha_i) + n_i,
\eeq
An unpolarized experiment is described by $y_i = T_{k_i} + n_i$.
Grouping the numbers into vectors, we can write any of these three expressions as 
a simple matrix equation 
\beq{LinearEq}
\y=\A\x + \n,
\eeq
where the matrix $\A$ encompasses all the relevant details of the observations.
For a pure polarization experiment as in \eq{sincosEq2}, 
$\A$ will contain only zeroes except for a single sine and cosine 
entry in each
row, in columns corresponding to the pixel observed.
More general observing strategies such as the beam-differencing of the MAP
Satellite
or modulated beams clearly retain the simple form of \eq{LinearEq}, merely with
a slightly more complicated (but known) $\A$-matrix.
For a well-designed experiment such as MAP, the system of equations 
(\ref{LinearEq}) is highly overdetermined,
and the estimate $\xt$ of the map triplet 
$\x$ given by the familiar equation \cite{mapmaking}
\beq{MinVarEq}
\xt=\W\y, \quad \W\equiv [\A^t\M\A]^{-1}\A^t\M
\eeq
is unbiased ($\expec{\xt} = \W\A\x +\W\expec{\n} = \x$ since $\W\A=\I$).
If $\M$ a reasonable approximation to $\N^{-1}$, then the
map noise $\W\n$ will have minimum variance to first order, with
covariance matrix $\NN\equiv\W\N\W^t\approx [\A^t\N^{-1}\A]^{-1}$.
Both the map triplet $\xt$ and its exact noise covariance matrix $\NN$ can be computed in 
$\sim N^3$ time, and even faster in many important cases
\cite{Wright96b,Wright96a,strategy,qmap3,Borrill,Revenu00,deBernardis00,Lange00,Hanany00,Szapudi00}. 
Note that the last $\P$-matrix is this noise covariance matrix,
\ie, $\P_{6\lmax-5}=\NN$. 

\subsubsection{The problem of missing modes}

In many cases, the inversion in \eq{MinVarEq} fails because 
the matrix to be inverted is singular.
Although there are typically much more measurements $y_i$ than
unknowns $x_i$ ($M\gg 3N$), symmetries or other properties of
the observing strategy often imply that $\A^t\x=\zero$ for 
certain vectors $\x$, \ie, that $\A$ 
is singular.
A ubiquitous example is lack of sensitivity to
the mean (monopole) in the map. 
Experiments measuring linear polarization without cross-linking
may be sensitive to only a certain linear combination of 
$Q$ and $U$, unable to recover the two separately.
As another example, the PIQUE experiment \cite{Hedman} measures 
only sums of $Q$-values $90^\circ$ apart on a circle in the sky, 
thereby losing information about modes taking values $(+1,-1,+1,-1)$ 
at four corners of a square.

All such problems can in principle be dealt with by regularizing the
inversion (see, \eg, the Appendix of \cite{cl}), which sets the unmeasurable modes
to zero in the final maps, and keeping track of 
which modes are missing during subsequent analysis of $\xt$.
In practice, however, it is often more convenient to 
eliminate this extra bookkeeping requirement by encoding the
corresponding information in the noise covariance matrix $\NN$ as
near-infinite noise for the missing modes. 
This ensures that the missing modes are given essentially zero weight 
in any subsequent analysis. 
This ``deconvolution'' technique is described in detail and tested numerically
in \cite{qmask}, and in practice corresponds to adding a matrix 
$\sigma^{-2}\I$ to the $[\A^t\M\A]$-term of  \eq{MinVarEq} before the inversion,
where $\sigma$ is about $10^2$ times the rms cosmological signal.
In summary, it allows {\it any} observed data set to be put in the
standard form of \eq{xDefEq}, described fully by 
the pair $(\xt,\NN)$ regardless of any missing modes.

\subsection{Measuring the power spectra with quadratic estimators}

In this section, we discuss how to measure the six power spectra
$C_\l^T$, $C_\l^E$, $C_\l^B$, $C_\l^{TE}$, $C_\l^{TB}$ and $C_\l^{EB}$
from the map triplet
$\xt$ of \eq{xDefEq}.
There are two basic approaches to this problem that ultimately 
give the same answer.

The first approach is to start by deconvolving the $Q$ and $U$ maps into 
$E$ and $B$ maps, and then use these as inputs to the power spectrum estimation.
Since $Q$ and $U$ depend linearly (but non-locally) on $E$ and $B$, this
can always be done with the deconvolution method of \cite{qmask}.
Incomplete sky coverage will simply be reflected by near-infinite noise 
in certain modes in the resulting noise covariance matrix. 
An advantage of taking this route is that Wiener-filtered $E$ and $B$ maps
can be plotted, whose spatial information may provide useful diagnostics
for foreground contamination and systematic errors.

The second approach, which we will adopt here, 
is to skip the intermediate step of $E$ and $B$ maps 
and measure the power spectra directly from $\x$.

\subsubsection{The definition of a quadratic estimator}

Our basic problem is to estimate the parameters $p_i$ in 
\eq{CdefEq} from the observed data set $\x$  
(we drop the tilde from \sec{mapSec} for simplicity).
Fortunately, this problem is mathematically identical to that for the 
unpolarized case, which has already been solved using so-called
quadratic estimators \cite{cl,BJK}. 
This class of methods is closely related to the maximum-likelihood 
method --- we return to this issue in \sec{DiscussionSec}.
A quadratic estimator $q_i$ is simply a quadratic function of the data vector $\x$,
so the most general case can be written as
\beq{qDefEq}
q_i\equiv\x^t\Q_i\x = \tr[\Q_i\x\x^t]
\eeq
where $\Q_i$ an arbitrary symmetric $3N\times 3N$-dimensional matrix. 
We will often find it convenient to group the parameters 
$p_i$ and the estimators $q_i$ into $\Nb$-dimensional 
vectors, denoted $\p$ and $\q$, where 
$\Nb=6\lmax-5$ is the number of bands.
The matrices $\Q_i$ should not be confused with the 
vector of stokes parameters $\Q$ from \eq{xDefEq}!

\subsubsection{The window function of a quadratic estimator}

Since $\Q_i$ can be any symmetric matrix, one can write down infinitely
many different quadratic estimators. 
Whether a given choice is useful or not depends on the 
mean and covariance of the vector $\q$. 
Equations\eqn{qDefEq} and\eqn{CdefEq} show that the mean of $\q$ is
\beqa{qMeanEq}
\expec{q_i}&=&\tr[\Q_i\C] = \sum_{i'} \tr[\Q_i\P_{i'}] p_{i'}\\
            &=& b + \sum_{P'=1}^6\sum_{\l'=2}^\lmax W^{\l P}_{\l'P'} C_{\l'}^{P'},
\eeqa
where $i=(\lmax-1)(P-1) + \l - 1$ is the parameter number corresponding to
polarization type $P$ and multipole $\l$, 
$b\equiv \tr[\Q_i\NN]$ is the contribution from experimental noise,
and
\beq{WindowDefEq}
W^{\l P}_{\l' P'}\equiv\tr[\Q_i\P_{i'}].
\eeq
These quantities can be viewed as a generalized form of window 
functions, since for a fixed $(P,\l)$, they show the expected contributions
to $q_i$ not only from different $\l$-values, but also from different
polarization types.

Ideally, we would be able to estimate $C_\l^P$ by
applying a quadratic estimator with the perfect window function
$W^{\l P}_{\l' P'}= \delta_{PP'}\delta_{\l\l'}$, but this is 
often impossible or undesirable with incomplete sky coverage,
shifting the aim to 
making the window functions narrow in both the
$\l$-direction and the $P$-direction. Minimizing such
unwanted mixing of different polarization types is one of the
key topics of this paper, and numerous examples of 
such window functions will be plotted in \sec{ResultsSec}.

The covariance matrix of $\q$ is 
\beq{qCovEq}
\M_{ij}\equiv\expec{\q\q^t}-\expec{\q}\expec{\q}^t
= 2\,\tr[\Q_i\C\Q_j\C]
\eeq
for the case where $\x$ is Gaussian, and it is 
clearly desirable to make it small in some sense.

\subsubsection{Quadratic estimators: specific examples}

It can be shown \cite{cl} that the quadratic estimator defined by
\beq{QdefEq}
\Q_i = {1\over 2}\norm_i\sum_j(\B)_{ij}\C^{-1}\P_j\C^{-1},
\eeq
distills all the cosmological information from $\x$ into the 
(normally much shorter) vector $\q$ if $\C$ is the true 
covariance matrix. Moreover, if $\C$ is a reasonable 
estimate of the true covariance matrix, say by computing 
it as in Appendix A using a prior power spectrum consistent with
the actual measurements, then the data compression step 
of going from $\x$ to $\q$ destroys information only to second order.
In \eq{QdefEq}, $\B$ is an arbitrary invertible matrix,
and the normalization constants $\norm_i$ are chosen so that 
all window functions sum to unity:
\beq{WindowNormEq}
\sum_{i'=1}^{\Nb-1}\tr[\Q_i\P_{i'}]=1.
\eeq
This means that we can interpret $q_i$ as measuring a weighted
average of our unknown parameters, the window giving the weights.
We will discuss a number of choices of $\B$ in \sec{ResultsSec}
that have various desirable properties.
$\B=\I$ gives minimal but correlated error bars.
$\B=\F^{-1}$ gives
beautiful Kronecker-delta window functions, corresponding 
to $\expec{\q}=\p$ at the price of anticorrelated and 
typically very large error bars, where
\beq{GaussFisherEq}
\F_{ij} = {1\over 2}\tr
\left[\C^{-1}{\partial\C\over\partial p_i}\C^{-1}{\partial\C\over\partial p_j}\right]
\eeq
is the so-called Fisher information matrix \cite{Fisher,karhunen} for the
case where the CMB fluctuations are Gaussian.
The intermediate choice $\B=\F^{-1/2}$ is normally a useful compromise \cite{texas96}, 
giving uncorrelated error bars and narrow window function with 
width $\Delta\l$ of order the inverse map size.
We will describe and test additional choices of $\B$ in 
\sec{DisentanglementSec} and \sec{CrossSec}.

\begin{table}[tb]
\begin{center}
{\sc Toy model specifications}
\begin{tabular}{lcrr}
Experiment	&Coverage	&FWHM	&Noise/pixel\\
\hline
COBE		&$b>20$		&$7^\circ$	&$30\mu K$\\
MAP		&$b>20$		&$13'$		&$71\mu K$\\
Planck		&$b>20$		&$8'$		&$ 8\mu K$\\
B2001		&$b>80$		&$12'$		&$16\mu K$\\
Circle		&$b=80$		&$12'$		&$16\mu K$\\
\end{tabular}
\end{center}
\end{table}

\subsubsection{Broadening the bands}
\label{BroadBandSec}

For CMB maps of small size where the window
function width $\Delta\l\gg 1$, it is unnecessary to oversample
the measured power with a separate parameter $p_i$ at each 
multipole $\l$. In such cases, it is useful to parametrize
the power spectrum as a staircase-shaped (piecewise constant) function,
with the parameters $p_i$ giving the heights of the various steps
\cite{BJK}. We will occasionally do this in \sec{ResultsSec}.


\section{Results}
\label{ResultsSec}

In this section, we will apply the quadratic estimator method to
a variety of fictitious data sets to quantify how 
experimental attributes such as sky coverage and sensitivity 
affect the ability to measure and separate the different power spectra.
We will also describe two ways in which the 
basic quadratic estimator technique can in some circumstances 
be improved for polarization applications.

\begin{figure}[tb]
\preskip
\centerline{\epsfxsize=9cm\epsffile{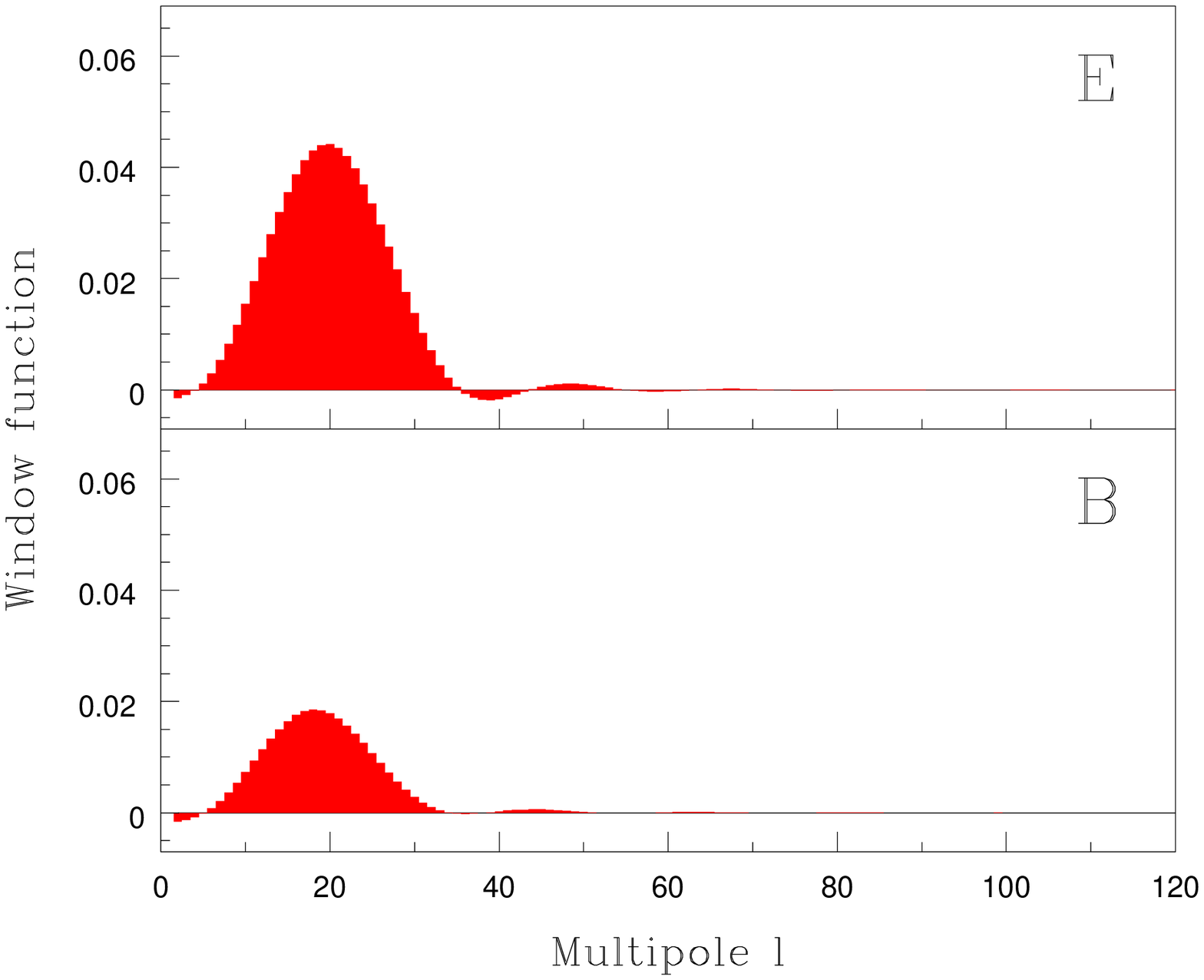}}
\postskip
\caption{\label{b2001WindowFig1}\footnotesize%
The window function corresponding to the B2001 measurement
of $\dT_\l^E$ for $\l=20$. 
Upper panel shows sensitivity to $E$-power 
(wanted) and 
lower panel shows sensitivity to $B$-power 
(unwanted --- what we call ``leakage'').
}
\end{figure}

\begin{figure}[tb]
\preskip
\centerline{\epsfxsize=9cm\epsffile{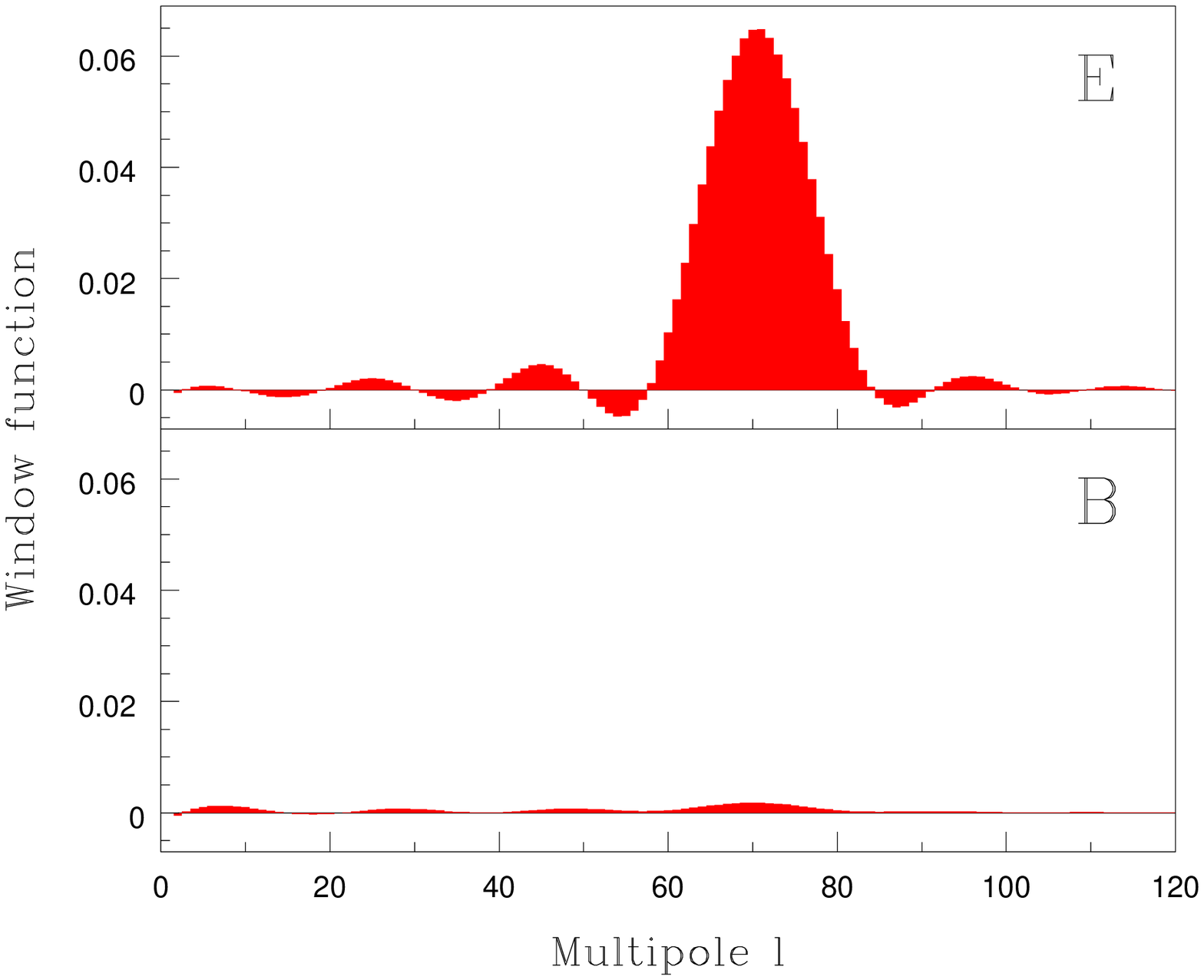}}
\postskip
\caption{\label{b2001WindowFig2}\footnotesize%
Same as \fig{b2001WindowFig1}, but for the B2001 $E$-measurement
aimed at $\l=70$.
}
\end{figure}

\subsection{Case studies}

Our five case studies are listed in Table 1.
The first three cover the northern Galactic cap $b>20^\circ$ with
successively higher sensitivity. The fourth covers a much smaller cap
$b>80^\circ$, and the fifth covers merely a one-dimensional
region: the circle defined by $b=80^\circ$.
These case studies are not intended to be 
accurate forecasts for the actual performance
of the experiments listed, but rather to span an interesting
range in sensitivity, sky coverage and map shape.
There are therefore numerous departures from realism.
For instance, the actual maps from COBE, MAP and Planck will of course
include the southern Galactic caps as well --- apart from reducing the error bars,
adding this reflection symmetry to the sky maps eliminates all leakage 
between even and odd $\l$-values \cite{cl}, preserving the
overall width of the window functions that we will present but 
giving them a jagged behavior where every other entry vanishes.
The actual map from the Boomerang 2001 (``B2001'') 
experiment will not be round and
will have non-uniform sensitivity.
We assume uncorrelated pixel noise for simplicity.
Most of the experimental sensitivities we have used are likely to be
slight underestimates, being based on a single frequency channel.

In all cases, we explicitly perform the various matrix computations
described in \sec{MethodSec}. The reason that this is numerically 
feasible within the scope of this paper is that the large angular scales
of interest here allow us to use larger and fewer pixels than the experimental
teams will employ in their actual data reduction. 

We will first study window functions to quantify how accurately 
$E$ and $B$ can be separated in various cases. We will then discuss 
measurement of the cross power spectrum and finally investigate how
accurately approximate error bars from the Fisher matrix formalism
match the results from our full numerical calculation.

\subsection{$E$ and $B$ window functions}
\label{EBsec}

We begin by quantifying the ability of $B2001$ and $MAP$ to separate 
$E$ and $B$ using $Q$- and $U$-maps. We pixelize our sky
patches using the equal-area icosahedron method \cite{icosahedron}
at resolution levels 35 and 7, 
respectively, corresponding to 361 B2001 pixels and 
561 MAP pixels\footnote{
We use the icosahedron pixelization since it has the roundest
(mainly hexagonal) pixels and is highly uniform.
Although we did not use it here, the HEALPIX package \cite{HEALPIX}
offers a useful alternative, allowing azimuthal symmetry to be exploited for 
saving computer time.}.
Since $Q$ and $U$ are measured for each pixel, the data vectors
$\x$ have twice these lengths. We use the method given
by \eq{QdefEq} with $\B=\F^{-1/2}$ unless otherwise specified.
We compute fiducial power spectra 
$C_\l^T$, $C_\l^E$ and $C_\l^{TE}$,
with the CMBFAST software
\cite{cmbfast} using cosmological parameters from the ``concordance''
model from \cite{concordance}, which provides a good fit to existing 
CMB and large scale structure data. We set $C_\l^{TB}=C_\l^{EB}=0$.
Although the true $B$-power spectrum may be close to zero,
we set $C_\l^B=C_\l^E$ in our fiducial model
since we wish to highlight geometrical effects. 
Since this prior is $E/B$-symmetric, any asymmetries between 
$E$ and $B$ in our resulting window functions and error bars
will be due to geometry alone. 
We eliminate sensitivity to offsets
by projecting out the mean (monopole) for the $T$, $Q$ and $U$ maps separately,
as described in the Appendix of \cite{cl}.

\begin{figure}[tb]
\preskip
\centerline{\epsfxsize=9cm\epsffile{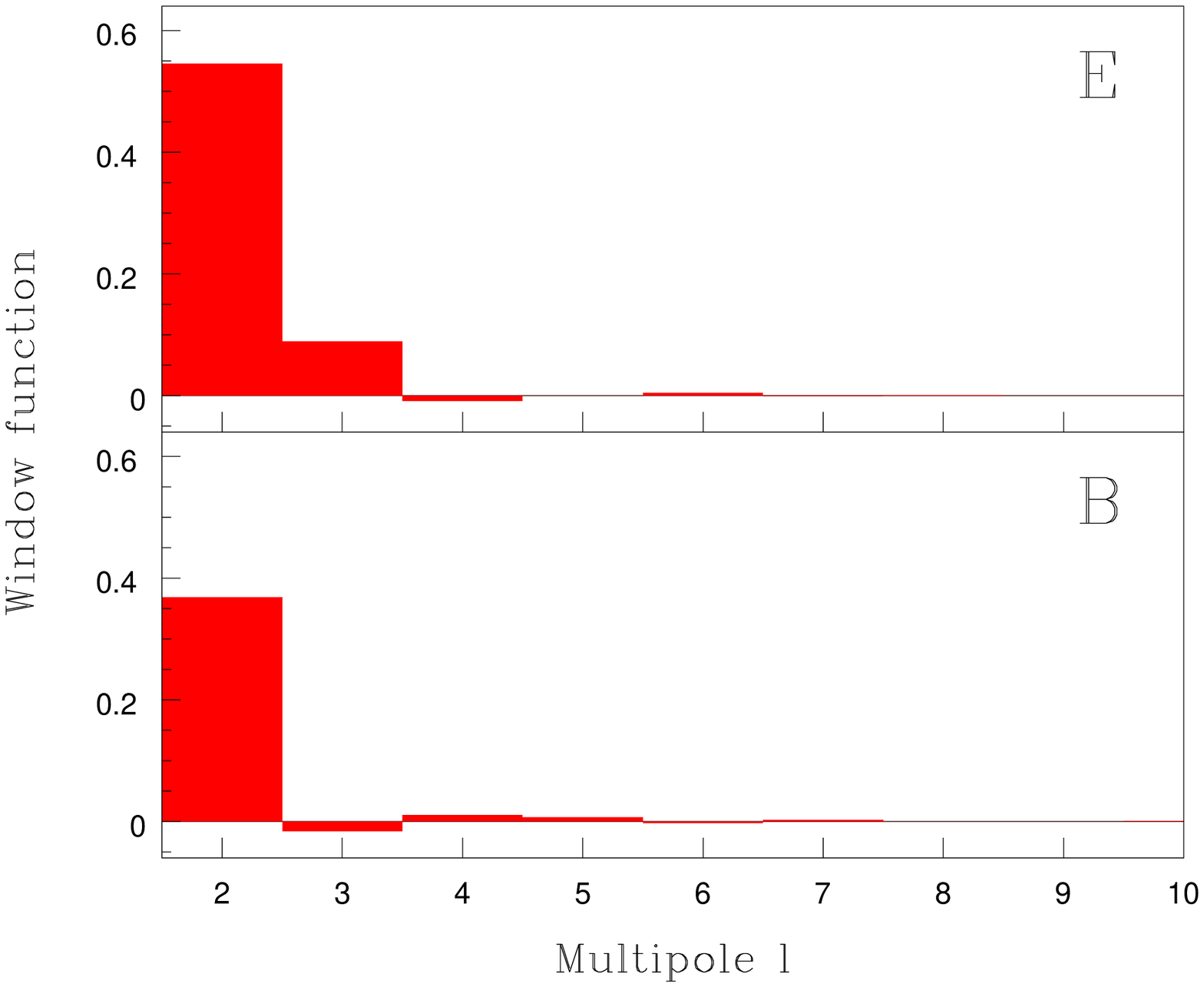}}
\postskip
\caption{\label{mapWindowFig1}\footnotesize%
Same as \fig{b2001WindowFig1}, but for the MAP $E$-measurement
aimed at $\l=2$.
}
\end{figure}

\begin{figure}[tb]
\preskip
\centerline{\epsfxsize=9cm\epsffile{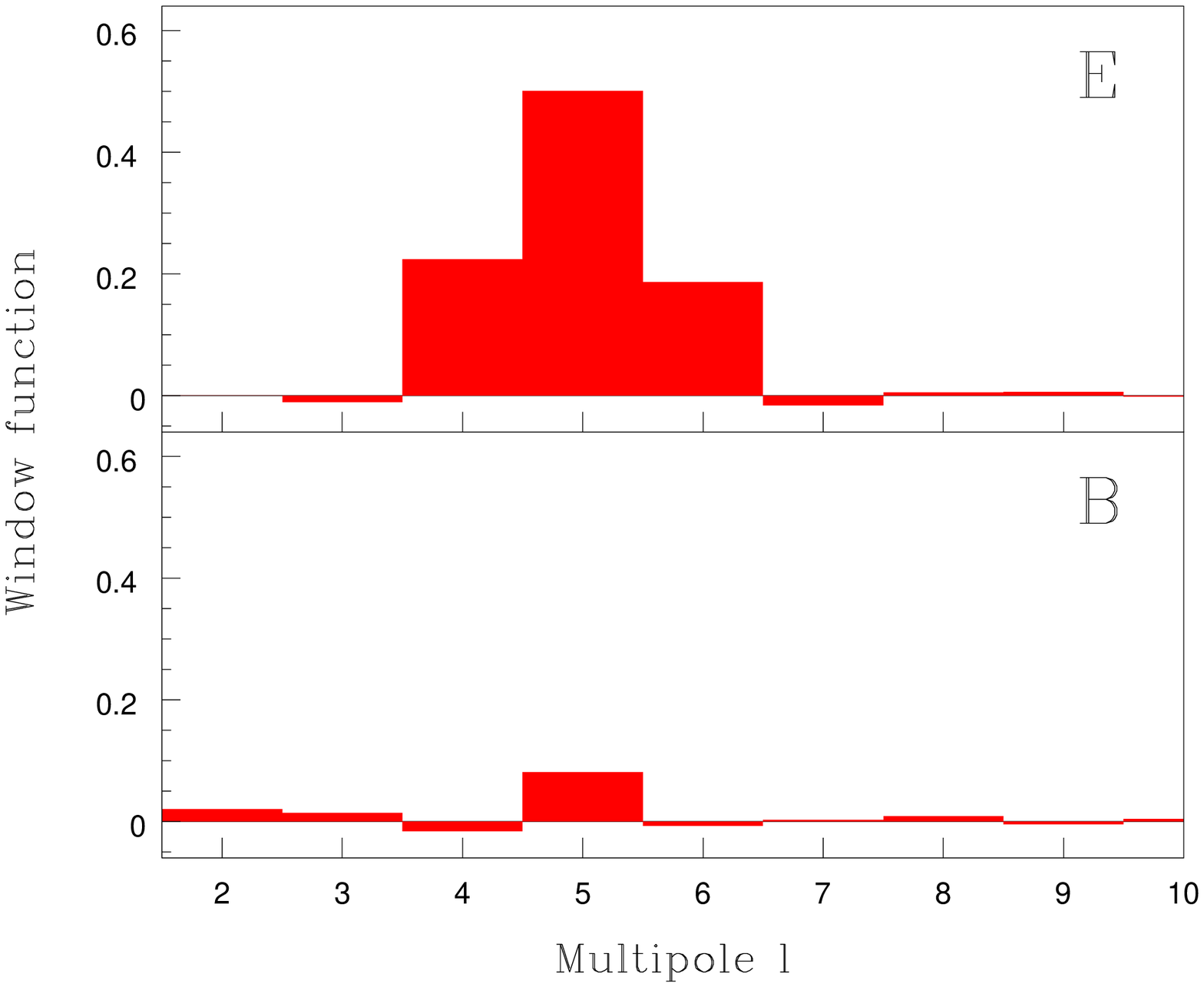}}
\postskip
\caption{\label{mapWindowFig2}\footnotesize%
Same as \fig{b2001WindowFig1}, but for the MAP $E$-measurement
aimed at $\l=5$.
}
\end{figure}

\begin{figure}[tb]
\preskip
\centerline{\epsfxsize=9cm\epsffile{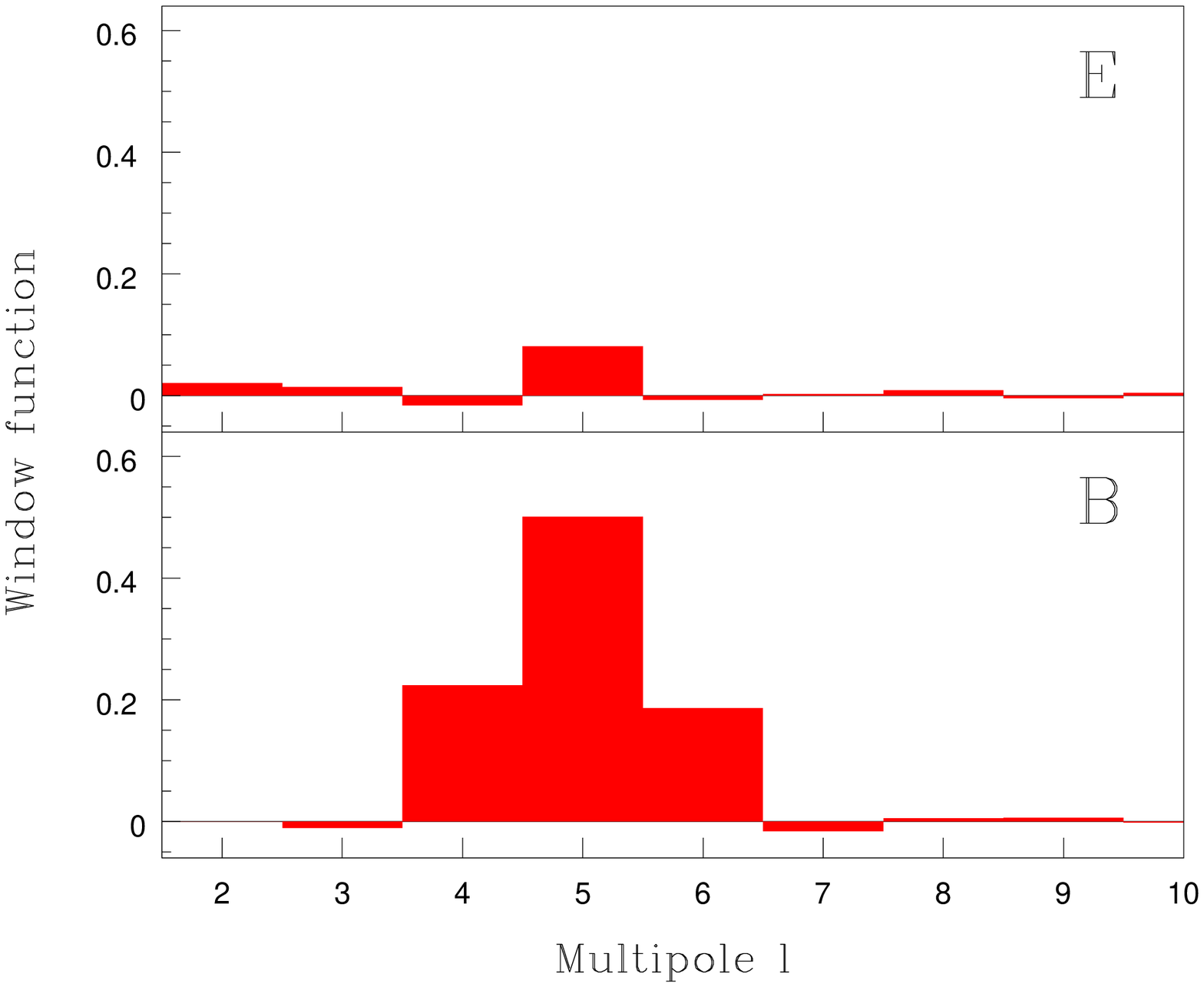}}
\postskip
\caption{\label{mapWindowFig3}\footnotesize%
Same as \fig{b2001WindowFig1}, but for the MAP measurement
of $\dT_\l^B$ for $\l=5$. 
}
\end{figure}

\subsubsection{Dependence on sky coverage and angular scale}

Figures~\ref{b2001WindowFig1}-\ref{mapWindowFig3} 
show a sequence of sample window functions for B2001 and MAP.
Note that it is possible for window functions to go slightly negative
for the decorrelation method $\B=\F^{-1/2}$ used here, whereas the 
method given by $\B=\I$ guarantees non-negative windows.
Just as for the unpolarized case, the
window function width $\Delta\l$ is seen to be fairly independent 
of the target multipole $\l$, essentially scaling as the inverse size
of the sky patch covered \cite{window}.

The amount of leakage of $B$-power into our $E$ estimate is quantified
by the lower panels in Figures~\ref{b2001WindowFig1}-\ref{mapWindowFig2}, 
and is seen to decrease as smaller scales are probed. 
\Fig{mapWindowFig3} targets $B$-polarization and looks like
\fig{mapWindowFig2} with the two panels swapped, thereby showing
that the leakage problems between $E$ and $B$ are quite symmetric.
The smaller the area under the unwanted half of the window function, the
better our method separates $E$ and $B$.
As a simple quantitative measure of this power leakage, let us therefore 
define a $2\times 2$ {\it leakage matrix} $\L^\l$ for each $\l$, given by
\beq{LdefEq}
\L^\l_{PP'}\equiv \sum_{\l'=2}^\lmax W^{\l P}_{\l' P'},
\eeq
where $P$ and $P'$ take the values $E,B$.
In other words, the four components of this leakage matrix
are the areas under the four histograms in  
Figure~\ref{mapWindowFig2} and~\ref{mapWindowFig3}.
If $\L^\l_{EE}=\L^\l_{BB}=1$ and $\L^\l_{EB}=\L^\l_{BE}=0$,
\ie, if $\L^\l=\I$, then there is on average no leakage at all between 
$E$ and $B$. 
For the simple case of complete sky coverage and uniform noise, all window
functions become Kronecker delta functions, 
$W^{\l P}_{\l' P'}=\delta_{PP'}\delta_{\l\l'}$,
and we verified that this happens numerically as a test of our software
(in practice, it works only when $\l,\l'$ are smaller than the scale corresponding to
the pixel separation, \ie, when the map is adequately oversampled).
This simple case thus gives the ideal case $\L^\l=\I$, but we will return 
in \sec{DisentanglementSec} to a method
producing this desirable result even for partial sky coverage.

\begin{figure}[tb]
\preskip
\centerline{\epsfxsize=9cm\epsffile{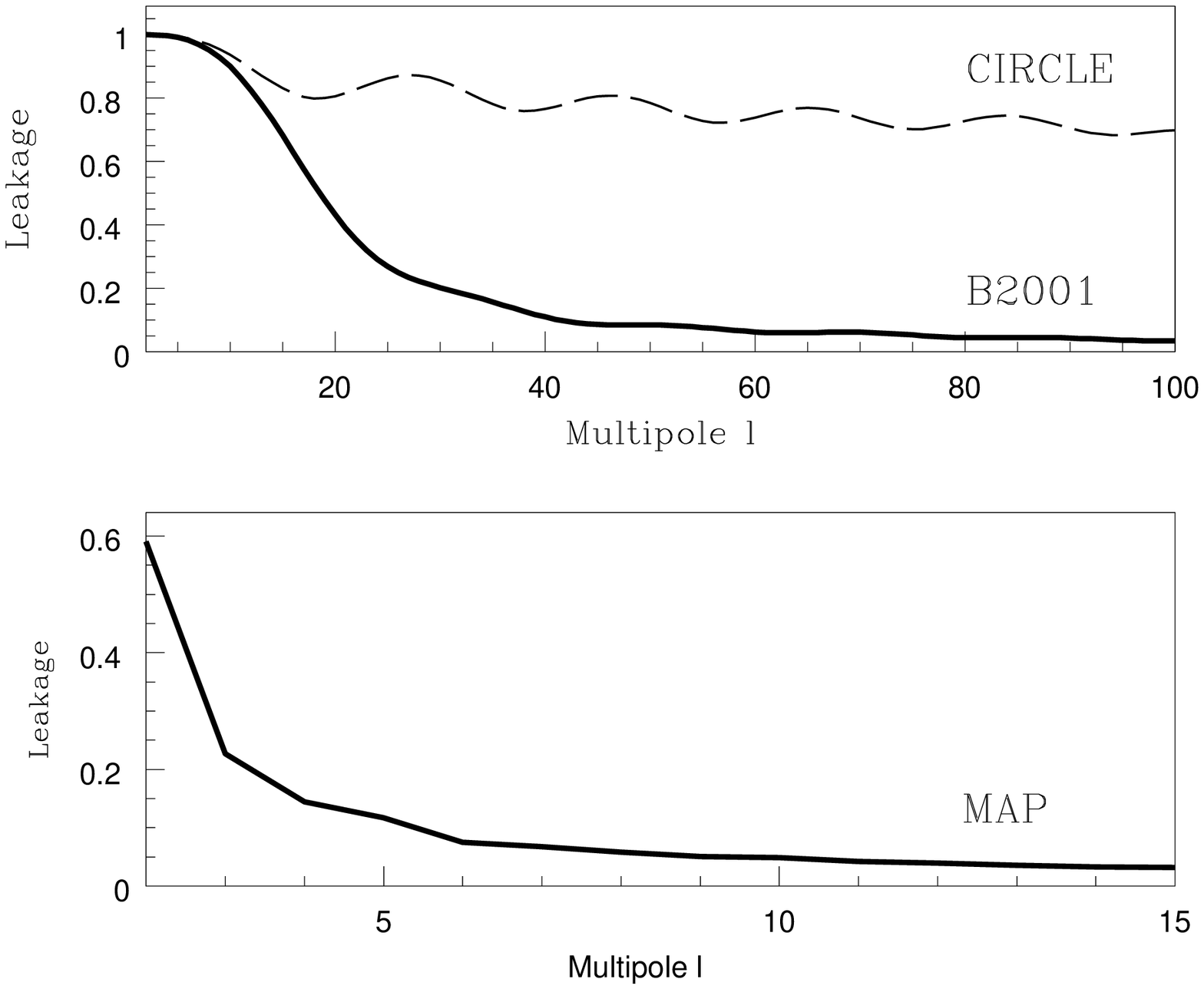}}
\postskip
\caption{\label{TripleLeakageFig}\footnotesize%
The amount of leakage of $B$-power 
into the $E$-power spectrum estimates is shown 
for the cases of B2001 (top panel, solid), 
CIRCLE (top panel, dashed) and MAP (bottom panel, solid).
These curves show the $B/E$ ratio ($\L^\l_{EB}/\L^\l_{EE}$)
for the $E$-estimates.
The corresponding curves for leakage in the reverse direction
(the $E/B$-ratio $\L^\l_{BE}/\L^\l_{BB}$ for the $B$-estimates)
are visually identical.
}
\end{figure}

To assess how the leakage depends on sky coverage and angular scale,
we plot the leakage for B2001, MAP and CIRCLE as a function of $\l$ in 
\fig{TripleLeakageFig}. Specifically, 
we plot the ratios of unwanted to wanted contributions, \ie,
$\L^\l_{EB}/\L^\l_{EE}$ and $\L^\l_{BE}/\L^\l_{BB}$.
These plots show three noteworthy results:
\begin{enumerate}
\item The situation for $E$ and $B$ is rather symmetric,
with essentially equal leakage from $B$ to $E$ as vice versa.
\item The leakage drops with $\l$.
\item The B2001 and MAP curves have roughly similar shape apart
from a scaling of the horizontal axis by a factor $\sim 7$,
corresponding to the map size ratio.
\end{enumerate}
Result 2 is expected since map boundary effects (incomplete sky coverage) 
are the reason that we cannot separate $E$ and $B$ perfectly
---  these boundary effects become less important 
as angular scales much smaller than the map are considered.
In the small-scale limit where sky curvature and discreteness of $\l$
become irrelevant, one would expect result 3 as well,
since there is no other $\l$-scale in the problem than 
the window function width $\Delta\l$, of order the
inverse size of the map. 
If $\theta$ denotes the diameter of our circular sky patches in radians,
then the FWHM window widths for B2001 and MAP are roughly fit by 
$\Delta\l\approx 5/\theta$, and 
the figures show that the leakage ratio drops below 
$15\%$ for $\l\simgt 2\Delta\l$.
(Things are different for the CIRCLE case, which we defer to 
\sec{CircleSec}.)

In conclusion, we have found that $E/B$ 
separation works well for $\l\gg\Delta\l$.

\begin{figure}[tb]
\preskip
\centerline{\epsfxsize=9cm\epsffile{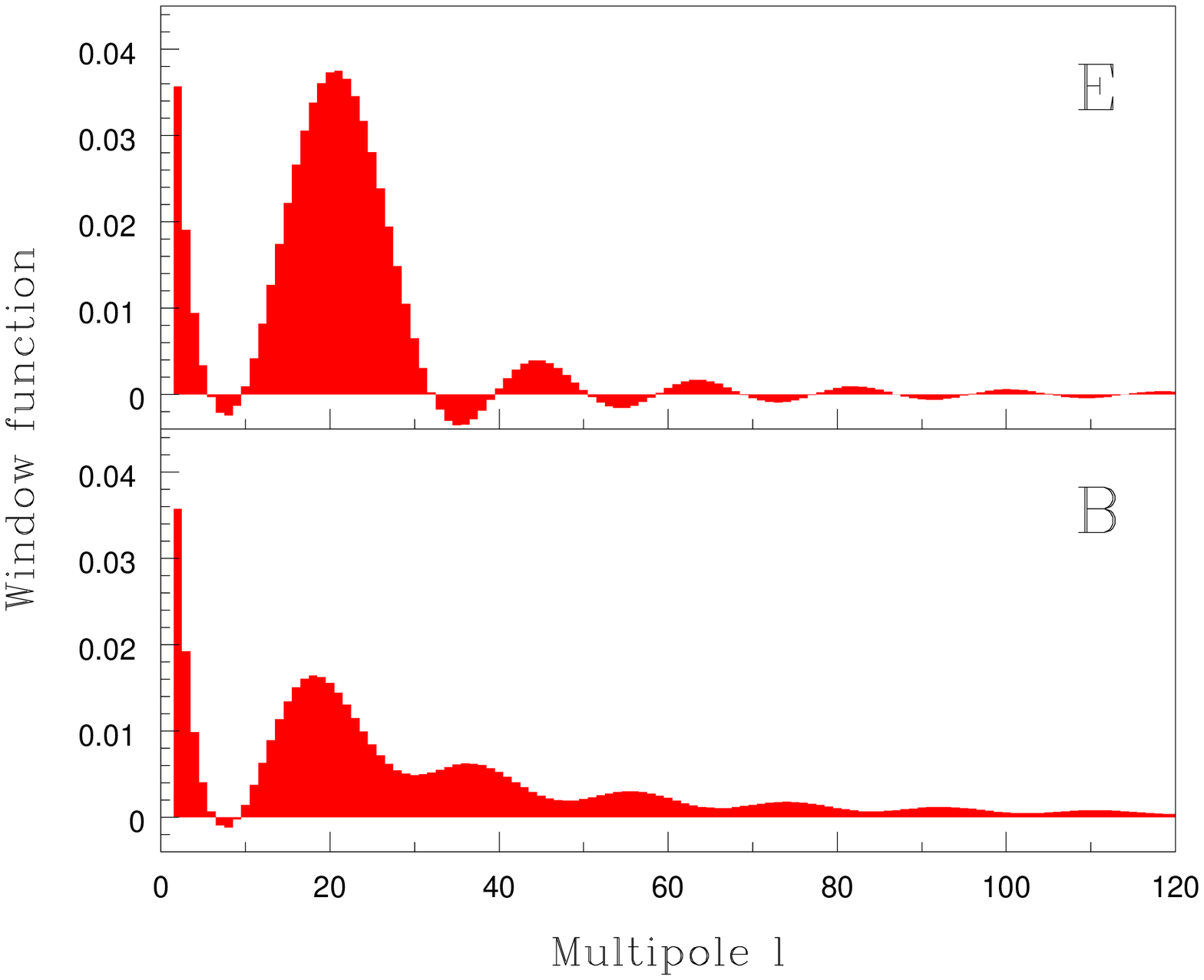}}
\postskip
\caption{\label{CircleWindowFig1}\footnotesize%
Same as \fig{b2001WindowFig1}, but for the CIRCLE $E$-measurement
aimed at $\l=20$.
}
\end{figure}

\begin{figure}[tb]
\preskip
\centerline{\epsfxsize=9cm\epsffile{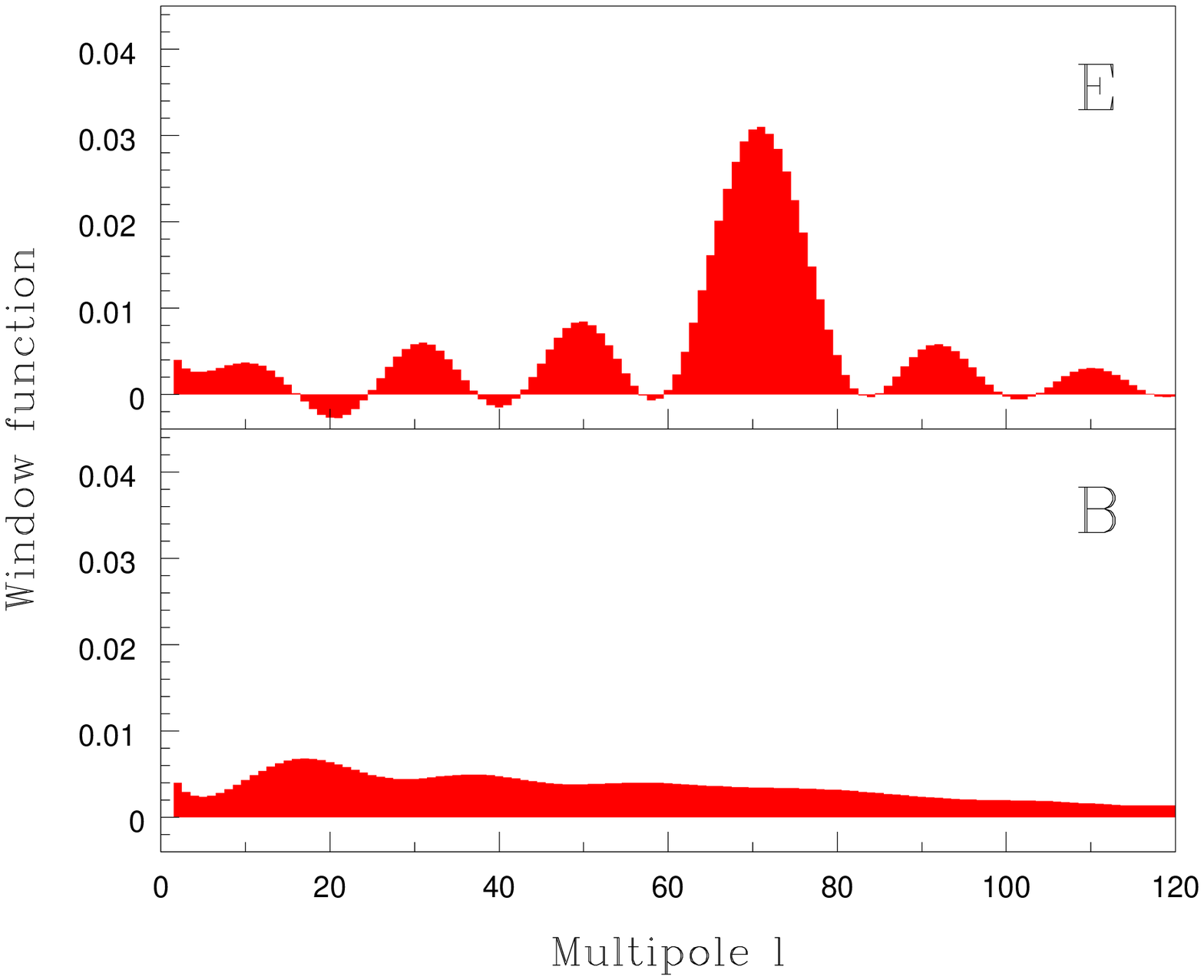}}
\postskip
\caption{\label{CircleWindowFig2}\footnotesize%
Same as \fig{b2001WindowFig1}, but for the CIRCLE $E$-measurement
aimed at $\l=70$.
}
\end{figure}


\subsubsection{Dependence on map shape}
\label{CircleSec}

Above we studied how leakage problems depend on map size.
To assess how they depend on map shape, we will now
compare two rather extreme examples: a disc and a circle.
The B2001 and CIRCLE maps have the same diameter and thus probe
comparable angular scales. However, whereas the B2001 map is 
truly two-dimensional, the CIRCLE map is essentially
one-dimensional, containing merely a single strip of pixels along the
circumference. The two current polarization experiments
POLAR \cite{Keating98} and PIQUE \cite{Hedman} both use ring-shaped
maps, and this important case has also
been extensively studied theoretically \cite{Zalda98}.

We pixelize our CIRCLE map with 360 equispaced pixels 
around the circle.
Sample CIRCLE window functions are shown in
Figures~\ref{CircleWindowFig1} and~\ref{CircleWindowFig2}.
It is seen that the CIRCLE windows are generally broader
(with larger $\Delta\l$) than their B2001 counterpart, 
which agrees with the well-known rule of thumb \cite{strategy} 
that the narrowest dimension of a map is the limiting factor.
We also see that whereas the B2001 leakage reduced on smaller scales,
things do not get correspondingly better for the CIRCLE case.
This explains the dashed curve in \fig{TripleLeakageFig}, 
which shows that substantial leakage persists even on small scales.
In \sec{DisentanglementSec}, we will describe how leakage can 
be further reduced.

In conclusion, we find that although ring maps do allow an interesting
degree of $E/B$-separation, a two-dimensional map works better
in this regard.

\subsubsection{Dependence on sensitivity}

Above we investigated how $E/B$-leakage was affected by map size and 
shape. To assess the effect of map sensitivity, we
compare our COBE, MAP and Planck examples. These have identical
sky coverage and pixelization, so the only difference is the
sensitivity per unit area which increases dramatically from 
COBE to MAP to Planck.
We refrain from plotting the three leakage curves, since they
look visually identical to the MAP curve in \fig{TripleLeakageFig}.
This means that the effect of sensitivity on $E/B$ separation is
negligible compared to the effect of sky coverage.
In other words, it depends mainly on geometry and only weakly on the
(sensitivity-dependent) details of the pixel weighting.

Generally, the quadratic estimator method strives to minimize
error bars by reducing leakage from multipoles
and polarization types with substantial power.
In a situation where sample variance is dominant, this 
tends to make windows slightly lopsided, with a wider
wing towards the direction where power decreases --- in most cases
towards the right. Conversely, in a situation where detector noise is
dominant, windows tend to be slightly lopsided in the opposite sense,
since noise power normally increases on smaller scales.


\subsection{Disentangling $E$ and $B$ better}
\label{DisentanglementSec}

Above we have seen that the $E$ and $B$ power spectra can be
fairly accurately separated on angular scales $\l\gg\Delta\l$
with the decorrelated quadratic estimator method.
Here we will argue that it is in some cases possible to do even 
better. 
Our motivation for pursuing this is that whereas broad windows in the
$\l$-direction are easy to interpret (corresponding simply to a 
smoothing of the power spectrum), the mixing of different polarization
types is rather annoying and complicates the interpretation of results. 

Each choice of $\B$ in \eq{QdefEq} corresponds to a different way of
plotting the results. Which is the best choice?

If the goal is to use the measured band power estimates
in $\p$ to constrain cosmological parameters, the choice is 
irrelevant. Any two methods using invertible $\B$-matrices
of course retain exactly the same cosmological information,
since it is possible to go back and forth between the corresponding 
two $\p$-vectors by multiplying by $\B_2\B_1^{-1}$ or $\B_1\B_2^{-1}$.
This means that the likelihood function for cosmological parameters
will be identical for the two methods. 

The choice of $\B$ is therefore mainly a matter of presenting the
power spectrum measurements in a clear and intuitive way.
Ideally, a method would have the following properties
\begin{enumerate} 
\item No leakage between $E$ and $B$
\item Narrow window functions
\item Uncorrelated error bars
\end{enumerate}
Unfortunately, these three properties are only achievable simultaneously
(without information loss) for the case of complete sky coverage.
As discussed in \cite{texas96}, the choice $\B=\F^{-1/2}$ in
\eq{QdefEq} achieves 3 and does a 
fairly good job on 2, giving window functions with the 
fundamental $\l$-resolution corresponding to the sky coverage.
Above we saw that this gives a leakage between $E$ and $B$ that
is substantial on scales corresponding to the size of the survey
and approaches zero on substantially smaller scales.


In contrast, the choice $\B=\F^{-1}$ in
\eq{QdefEq} achieves 1 and 2 perfectly, but 
at the cost of producing measurements that are difficult
to interpret because the error bars are typically 
anticorrelated and huge \cite{texas96}.
However, these problems are to a large extent caused by
eliminating the rather benign leakage between different
$\l$-values. We will now describe a less aggressive
method that merely targets the $E/B$-leakage.
Specifically, let us insist that the leakage matrices $\L^\l=\I$
for all $\l$.

Let us define the {\it disentangling method} as the one given
by \eq{QdefEq} with the $\B$-matrix chosen as 
\beq{DisentangledBeq}
\B_{ii'} = \B_{\P\l\P'\l'} 
\equiv\sum_{P''}(\L_\l^{-1})_{PP''} \F^{-1/2}_{\P''\l\P'\l'}.
\eeq
Here we have once again combined $P$ and $\l$ into 
a single index $i = (\lmax-1)(P-1) + \l - 1$.
It is easy to show that this method gives ideal leakage matrices
$\L^\l=\I$ if the leakage matrices in \eq{DisentangledBeq}
were computed using $\B=\F^{-1/2}$. This means that
the unwanted half of the window function averages to zero.
A similar scheme was found to work well for disentangling
three types of power in a galaxy redshift survey \cite{pscz}.

\begin{figure}[tb]
\preskip
\centerline{\epsfxsize=9cm\epsffile{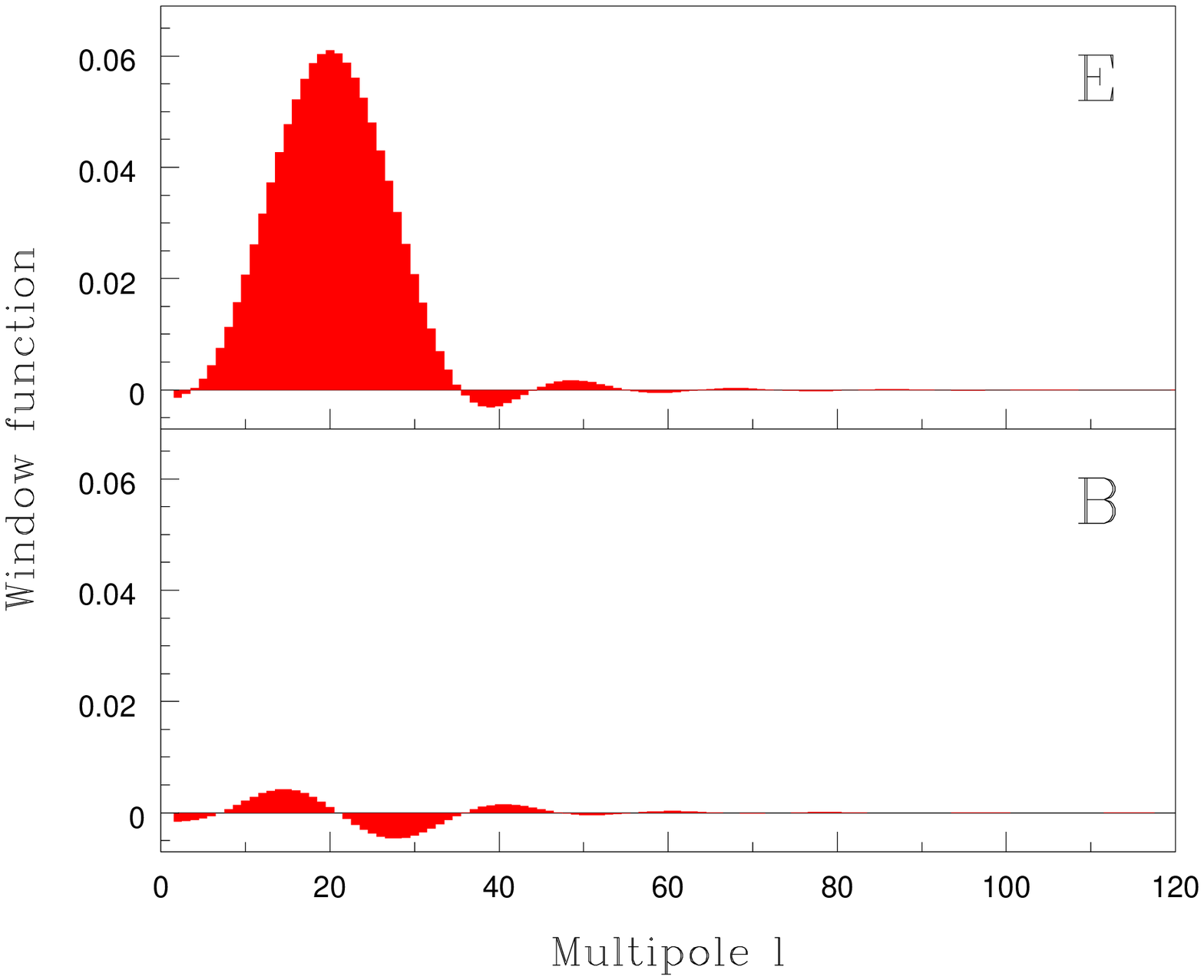}}
\postskip
\caption{\label{DisentangledWindowFig}\footnotesize%
Same as \fig{b2001WindowFig1}, but after applying our disentanglement method.
}
\end{figure}

\begin{figure}[tb]
\preskip
\centerline{\epsfxsize=9cm\epsffile{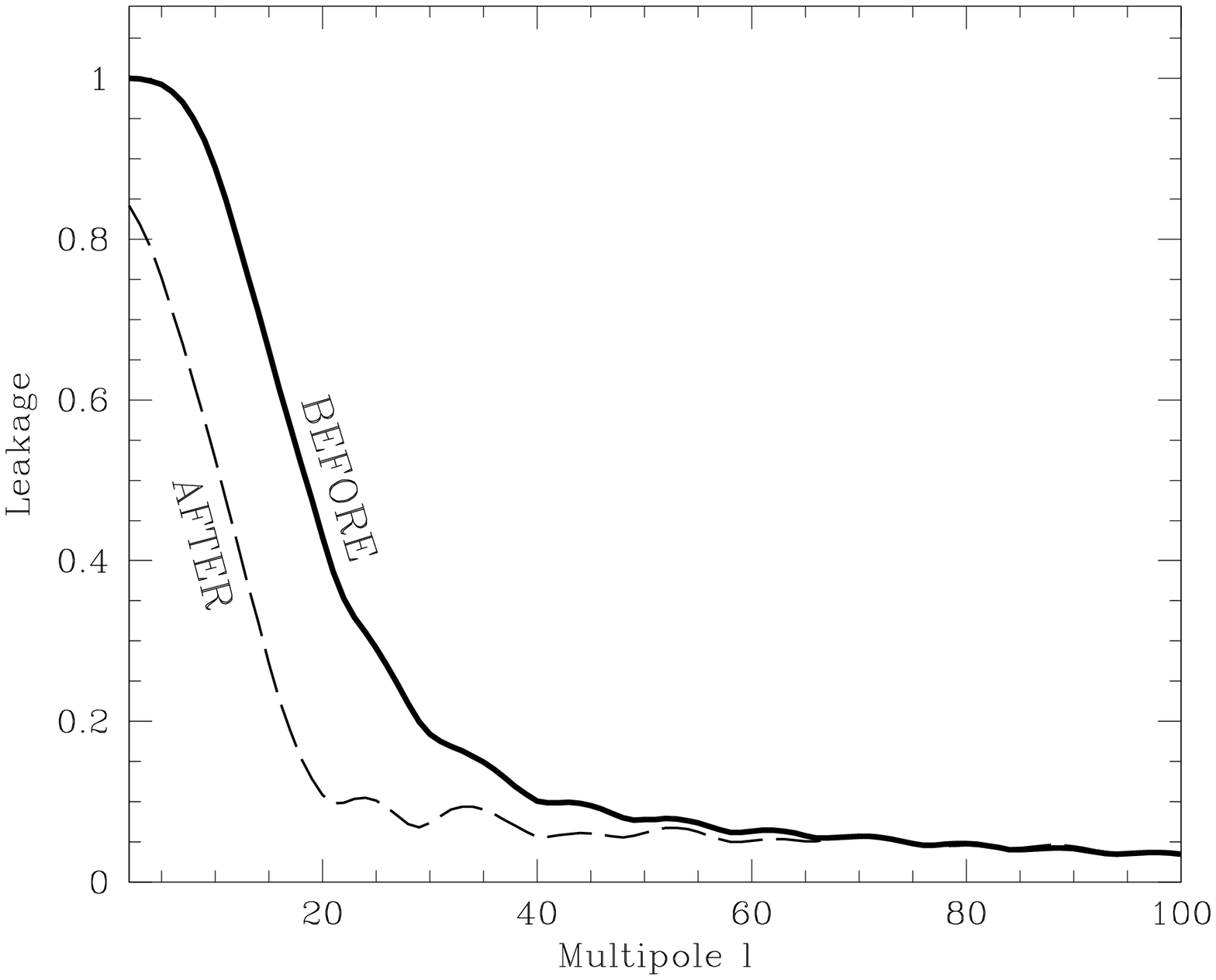}}
\postskip
\caption{\label{DisentanglementLeakageFig}\footnotesize
Curves showing leakage of $B$-power into estimates of $E$
are plotted for B2001 before (solid) and after (dashed) applying
out disentanglement method. In this plot, leakage has been
computed by taking absolute values of the window functions
--- otherwise the dashed curve would be identically zero.
For example, the value of the dashed curve at $\l=20$ can 
be interpreted simply as 
the ratio of shaded area in the two panels of 
\fig{DisentangledWindowFig}.
}
\end{figure}

We test this method for our B2001 example, and a sample disentangled
window function is shown in \fig{DisentangledWindowFig}.
Comparing this with \fig{b2001WindowFig1}, which targeted the 
same multiple, we see that the leakage (lower panel) has been 
substantially reduced and oscillates around zero.
On very large angular scales $\l\sim\Delta\l$, this undesirable
half of the window function remains substantial even though it
by construction averages to zero.
On small scales, however, the unwanted part of the window function
is found to be consistently near zero, not merely on average.
This is because both the desirable and the undesirable halves
of the initial window function before disentanglement 
have essentially the same shape, so that our disentanglement
process will cancel them out almost completely.
To quantify how well the this process works, 
\fig{DisentanglementLeakageFig} shows leakage curves with
the leakage matrix redefined with absolute values:
\beq{DdefEq}
\Lt^\l_{PP'}\equiv \sum_{\l'=2}^\lmax \left|W^{\l P}_{\l' P'}\right|,
\eeq
(Without taking absolute values, the disentanglement method
would by construction be diagnosed with zero leakage.)

Although this method is likely to be adequate for practical applications,
it is possible to disentangle $E$ and $B$ 
still better if necessary, at the price of larger
error bars.
The Fisher matrix $\F$ generally becomes singular if 
the bands used are much narrower than $\Delta\l$.
If the sky coverage is large enough that $\Delta\l$ is narrower
than features of interest in the power spectra, is it convenient 
to choose bands of width around $\Delta\l$ instead of width one
(solving for each multipole separately). Since this produces
an invertible Fisher matrix, perfect disentanglement is achievable
by setting $\B=\F^{-1}$ in \eq{QdefEq}, giving only a modest increase in
error bars and rather slight anticorrelations between neighboring bands.
The resulting measurements can then be made uncorrelated separately
for $E$ and $B$, by multiplying by the inverse square roots of 
the corresponding two covariance matrices, thereby broadening the
$\l$-windows back to their natural widths.

\subsection{Measuring the cross power spectrum}
\label{CrossSec}

\begin{figure}[tb]
\preskip
\centerline{\epsfxsize=9cm\epsffile{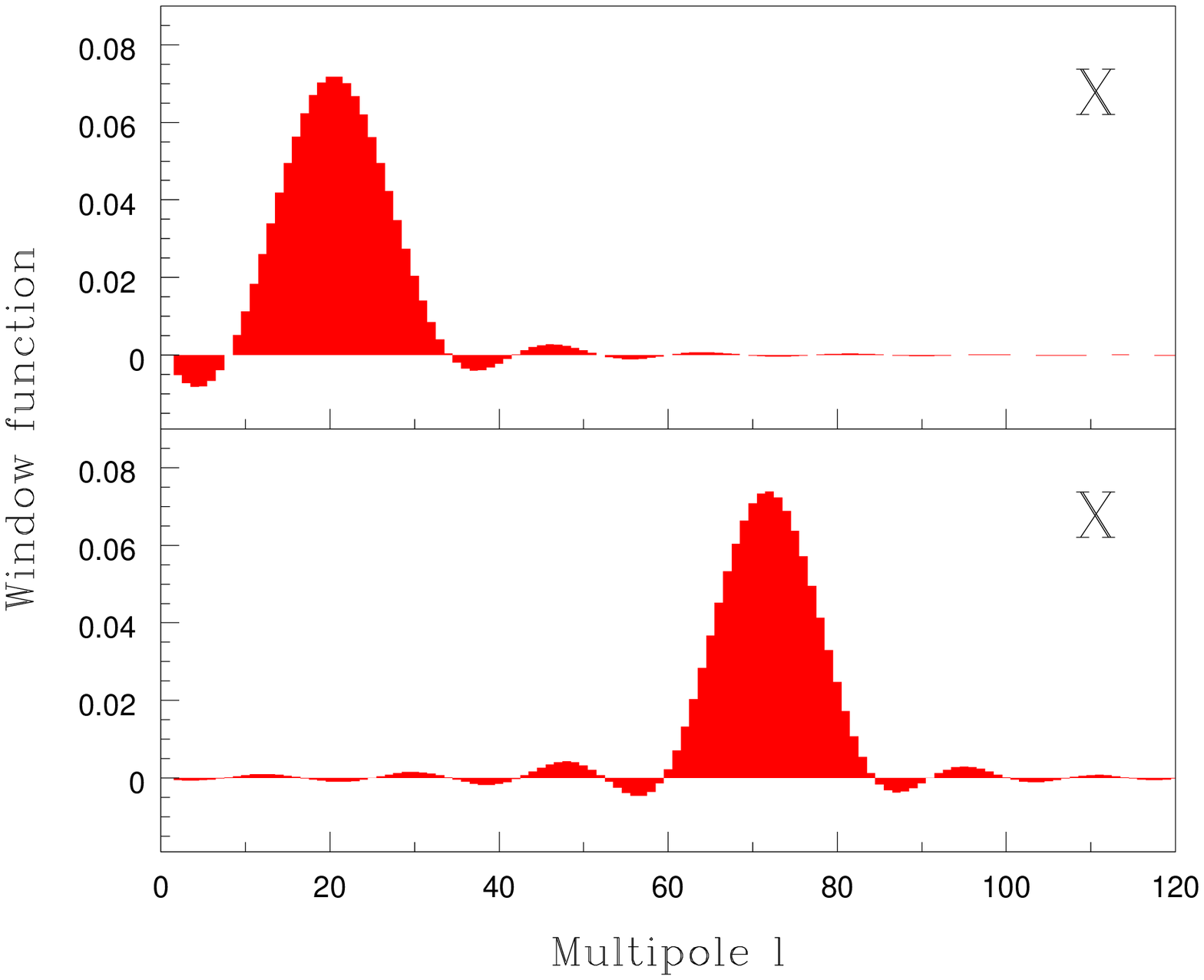}}
\postskip
\caption{\label{CrossWindowFig}\footnotesize%
Window functions for measuring the cross power spectrum
are shown for the B2001 case, targeting $\l=20$ (top) and $\l=70$ (bottom).
}
\end{figure}

The three cross power spectra $C_\l^{TE}$, $C_\l^{TB}$ and $C_\l^{EB}$,
which we will denote $C_\l^X$, $C_\l^Y$ and $C_\l^Z$ for brevity,
can be measured using
the basic decorrelated quadratic estimators given by 
\eq{QdefEq} without any modification. 
However, as we will now discuss, 
this is not necessarily the most desirable approach.

To measure the cross power spectrum $C_\l^X$,
our data vector $\x$ must contain both unpolarized and polarized
measurements as in \eq{xDefEq}. These may be either from a single
experiment or from separate unpolarized and polarized ones.
As reviewed in Appendix A, this gives a covariance matrix of
the form
\beq{FullCorrEq}
\C\equiv\expec{\x\x^t} =  
\left(\begin{tabular}{ccc}
$\expec{\T\T^t}$	&$\expec{\T\Q^t}$	&$\expec{\T\U^t}$\\[2pt]
$\expec{\Q\T^t}$	&$\expec{\Q\Q^t}$	&$\expec{\Q\U^t}$\\[2pt]
$\expec{\U\T^t}$	&$\expec{\U\Q^t}$	&$\expec{\U\U^t}$\\[2pt]
\end{tabular}\right)
\eeq
which generically has non-vanishing elements in all the off-diagonal 
blocks. 
When we try to measure one of the parameters $p_i$ corresponding to
the cross power spectrum $C_\l^X$, the 
matrix $\P_i=\partial\C/\partial p_i$ will vanish
except in the $T-Q$ and $T-U$ cross-correlation blocks, since these are
the only ones that depend on $C_\l^X$.
However, since $\C$ is a full matrix, 
the quadratic estimator $\Q_i\propto\C^{-1}\P_i\C^{-1}$ 
(as well as decorrelated or disentangled variants thereof) will also be
a full matrix, without any vanishing blocks.
This means that the estimates $q_i$ of the cross power spectrum will
involve not only data combinations like $T_j Q_k$ and $T_j U_k$, 
but also terms like $T_j T_k$ and $Q_j Q_k$. In other words,
the measured cross-correlation involves, among other things, the correlation
of the temperature map with itself!
The same peculiarity applies to the maximum-likelihood method, 
which is simply an iterated version of the quadratic method.

We will now give a simple example illustrating why this happens, as well as
argue that it is avoidable and sometimes undesirable.

\subsubsection{A complicated way to measure correlation}

As an illustrative toy model, let us temporarily assume that our data vector $\x$ 
consists of merely two numbers, the measurements of $T$ and $E$ 
for a given multipole $(\l,m)$ extracted from an all-sky map.
Consider the simple case where $C_\l^T=C_\l^E=1$ --- the general case 
will follow from our result by a trivial scaling.
This means that the data covariance matrix takes the form
\beq{ToyCeq}
\C\equiv\expec{\x\x^t}   
= \left(\begin{tabular}{cc}
$\expec{C_\l^T}$	&$\expec{C_\l^X}$\\[2pt]
$\expec{C_\l^X}$	&$\expec{C_\l^E}$
\end{tabular}\right)
= \left(\begin{tabular}{cc}
$1$	&$r$\\[2pt]
$r$	&$1$\\[2pt]
\end{tabular}\right),
\eeq
where $r\equiv C_\l^X/[C_\l^T C_\l^E]^{1/2}$ is the dimensionless
correlation coefficient between $T$ and $E$.
We have only three parameters to measure in our toy example, 
$\p=(C_\l^T,C_\l^E,C_\l^X)$,
so the matrices $\P_i=\partial\C/\partial p_i$ are 
\beq{ToyPeq} 
\P_1 = \left(\begin{tabular}{cc}
$1$	&$0$\\[2pt]
$0$	&$0$\\[2pt]
\end{tabular}\right),
\quad
\P_2 = \left(\begin{tabular}{cc}
$0$	&$0$\\[2pt]
$0$	&$1$\\[2pt]
\end{tabular}\right),
\quad
\P_3 = \left(\begin{tabular}{cc}
$0$	&$1$\\[2pt]
$1$	&$0$\\[2pt]
\end{tabular}\right).
\eeq
Substituting this and \eq{ToyCeq} into \eq{GaussFisherEq} gives 
the Fisher matrix
\beq{ToyFeq} 
\F = {1\over(1-r^2)^2}
\left(\begin{tabular}{ccc}
${1\over 2}$	&${r^2\over 2}$	&$-r$\\[2pt]
${r^2\over 2}$	&${1\over 2}$	&$-r$\\[2pt]
$-r$	&$-r$	&$1+r^2$\\[2pt]
\end{tabular}\right),
\eeq
with inverse
\beq{ToyFinvEq}
\F^{-1} = {1\over 2}
\left(\begin{tabular}{ccc}
$1$	&$r^2$	&$r$\\[2pt]
$r^2$	&$1$	&$r$\\[2pt]
$r$	&$r$	&${1+r^2\over 2}$\\[2pt]
\end{tabular}\right).
\eeq
Substituting the above equations into \eq{QdefEq}
and using the method given by $\B=\F^{-1}$, the
resulting unbiased estimators take the
simple form 
\beq{ToyQeq} 
\Q_1 = \left(\begin{tabular}{cc}
$1$	&$0$\\[2pt]
$0$	&$0$\\[2pt]
\end{tabular}\right),
\quad
\Q_2 = \left(\begin{tabular}{cc}
$0$	&$0$\\[2pt]
$0$	&$1$\\[2pt]
\end{tabular}\right),
\quad
\Q_3 = {1\over 2}\left(\begin{tabular}{cc}
$0$	&$1$\\[2pt]
$1$	&$0$\\[2pt]
\end{tabular}\right).
\eeq
This is not surprising, since no other estimators could
possibly be unbiased ($\expec{\q}=\p$) when only one spherical 
harmonic mode
is used. As soon as more modes are available, however, things can 
(and generally do) get
more complicated. Consider, the case where $\x$ consists of four
rather than two measurements for our multipole, 
--- temperature measurements $T_1$ and $T_2$ 
and $E$-polarization measurements $E_1$ and $E_2$.
The obvious guess would be that and estimator of 
$C_\l^X$ should involve only cross terms ($T_1 E_1$, $T_1 E_2$
and $T_2 E_2$). However, terms such as
$T_1^2-T_2^2$ and $E_1^2-E_2^2$ will have a vanishing average, and
can therefore be added to any estimator 
without biasing it. As we saw above, precisely this generically 
happens in our minimum-variance estimator, since it 
helps reduce the estimator variance if our four measurements have 
different noise variances.

This can also be understood as follows. Repeating the derivation 
of \eq{ToyQeq} when there are $n$ measurement of $T$ and $E$ for
our mode, perhaps with different variances, 
shows that the zeroes in \eq{ToyQeq} will be replaced by
$n\times n$ blocks that generally not vanish --- merely their
traces will.

\subsubsection{A simpler way}

Apart from being surprisingly complicated to compute and interpret,
the basic quadratic estimators of \eq{qDefEq} also have 
drawback related to systematic errors. If the estimated cross-power spectrum 
involves components of the unpolarized and polarized power spectra
that are supposed to cancel out to reduce the variance, there is a risk
that systematic errors from these two 
autocorrelations propagate into and contaminate 
the cross-correlation measurements.
For this reason, an interesting alternative is to use an estimator 
for the cross-power spectrum that does not have this funny property, 
\ie, that contains only cross-terms. A simple way to construct such an
estimator is to use \eq{QdefEq} with the fiducial power spectrum
$C_\l^X$ set equal to zero. This will make $\C$ and $\C^{-1}$ block diagonal,
so the $\Q$-matrices of \eq{QdefEq} acquire the
same block structure as the $\P$-matrices. 
The estimators of $C_\l^X$ therefore involve only cross terms.

\begin{figure}[tb]
\preskip
\centerline{\epsfxsize=9cm\epsffile{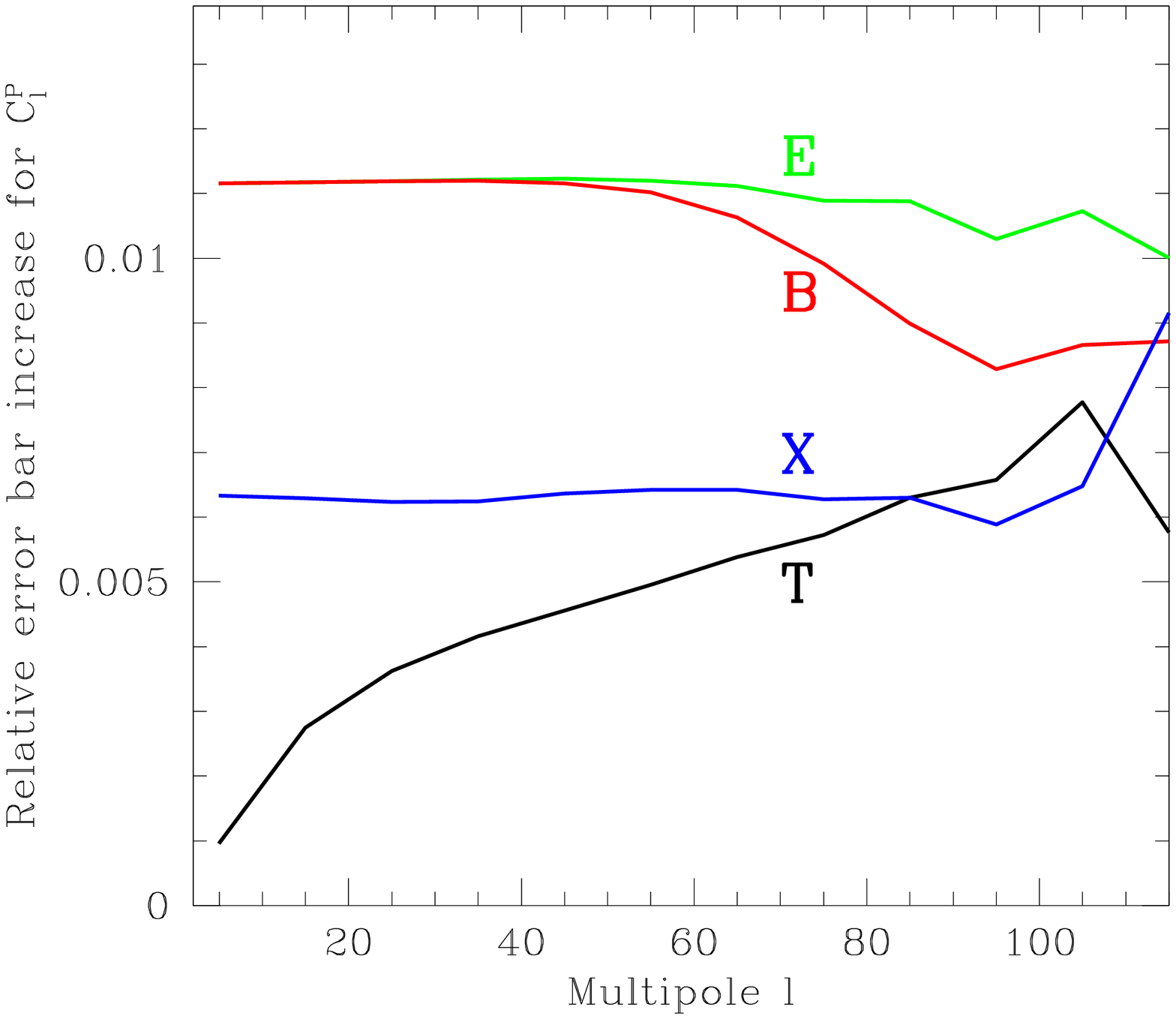}}
\postskip
\caption{\label{CrossErrorFig}\footnotesize%
The low cost of simplicity. The curves show the relative 
increase of the error bars on the four power spectra when
the fiducial model is replaced by one with $\C_\l^X=0$,
thereby eliminating potential systematic
errors as described in the text.
}
\end{figure}

It is noteworthy that this issue applies not only to estimation of 
$C_\l^X$, but to measuring $C_\l^T$, $C_\l^E$ and $C_\l^B$ as well.
If the fiducial power spectrum has $C_\l^X\ne 0$, then estimators of
all four power spectra will involve using combinations of all 
three maps $(\T,\Q,\U)$, so setting $C_\l^X\ne 0$ when using 
\eq{QdefEq} is an interesting simplifying option for all power
spectrum estimation. We implicitly did so in \sec{EBsec} by 
ignoring the $T$-map.

\subsubsection{Which method is better?}

The price we must pay for this simplification is a slight increase
in error bars. This is quantified in \fig{CrossErrorFig} 
for the B2001 case.
We computed the error bars on the four standard power power spectra
for the unbiased method with both weighting schemes 
(assuming the true $C_\l^X$ for our concordance model
and assuming $C_\l^X=0$)\footnote{
Specifically, we use broad $\l$-bins as described
in \sec{BroadBandSec} with $\Delta\l=10$ and use 
the method with $\B=\F^{-1}$ in \eq{QdefEq} 
to ensure that we are comparing apples with apples, \ie, 
to ensure that we are comparing error bars for 
measurements with identical window functions.
}. 
The figure shows the ratios (minus one), 
and illustrates that the simplification 
typically comes at a very low cost --- an error bar increase of 
order a percent. 
In light of this and the potential peril of systematic errors, 
the simpler method appears preferable in practice. 
Returning to the most general case of estimating six joint power spectra,
it is thus prudent to set all three cross power spectra to zero in the
fiducial model: $C_\l^X=C_\l^Y=C_\l^Z=0$.

\subsection{Error bars}

Up until now, we have focused on the issue of window functions.
Let us now turn to the complementary issue of error bars.
A large number of papers, for instance
\cite{Zalda97,parameters2,foregpars,JaffePursuer,Kinney98,Knox99,Prunet00,Popa00,Bucher00}, 
have made forecasts for how accurately upcoming polarization
experiments can constrain cosmological parameters.
These estimates all assumed that the accuracy of the 
recovered polarized power spectra $(C_\l^T,C_\l^E,C_\l^B,C_\l^X)$
would be given by the $4\times 4$ covariance matrix \cite{ZSS97}
\beq{PowerCovEq}
\M_\l\approx {2\fsky^{-1}\over 2\l+1}
\left(\begin{tabular}{cccc}
$T_\l^2$	&$X_\l^2$ & 0	&$T_\l X_\l$\\[2pt]
$X_\l^2$	&$E_\l^2$ & 0	&$E_\l X_\l$\\[2pt]
$0$		&$0$		&$B_\l^2$	&$0$\\[2pt]
$T_\l X_\l$	&$E_\l X_\l$	& 0 	&${1 \over 2}[T_\l E_\l+X_\l^2]$
\end{tabular}\right).
\eeq
where $\fsky$ is the fraction of the sky observed
and 
\beq{NoiseEq}
P_\l\equiv C_\l^P + (w^{P})^{-1} e^{\theta^2\l(\l+1)}, 
\quad P=T,E,B,X,
\eeq
if the experimental beam is Gaussian with width $\theta$ in radians 
(the full-width-half-maximum is given by FWHM$=\sqrt{8\ln 2}\>\theta$).
Here the sensitivity measure $1/w^{P}$ is defined as 
\cite{Knox95} the
noise variance per pixel times the pixel area in steradians for $P=T,E,B$.
For the case of the $EB$ cross power spectrum,
$w^X=(w^T w^E)^{1/2}$.
\Eq{NoiseEq} can be interpreted as simply giving 
the total power from CMB (first term) and detector noise (second term).
For our examples, we take 
$w^T = w^E = w^B = 1/($FWHM$\>\sigma)^2$,
where FWHM and $\sigma$ are the resolution and noise values
listed in Table 1.

The approximation 
of \eq{PowerCovEq} has been shown to be exact for
the special case of complete sky coverage $(\fsky=1)$ \cite{Zalda97}, 
and the same result follows
from the Fisher information matrix formalism \cite{foregpars}.
The factor $(2\l+1)\fsky$ can be interpreted as the effective number of uncorrelated
modes per multipole\footnote{
\label{fskyFootnote}
When the sky coverage $\fsky<1$, certain multipoles become correlated
\cite{window}. 
This reduces the effective number of uncorrelated modes
by a factor $\fsky^{-1}$, thereby increasing the sample 
variance on power measurements by the same factor
\cite{ScottCosmicVar,Knox95}. It also smears out sharp 
features in the power spectrum by an amount 
$\Delta\l$ comparable to the inverse size of the sky patch
in its narrowest direction \cite{strategy} and mixes $E$ and $B$ power
as discussed in the previous sections.
}, 
and the other factor as giving the covariance per mode.

\begin{figure}[tb]
\preskip
\centerline{\epsfxsize=9cm\epsffile{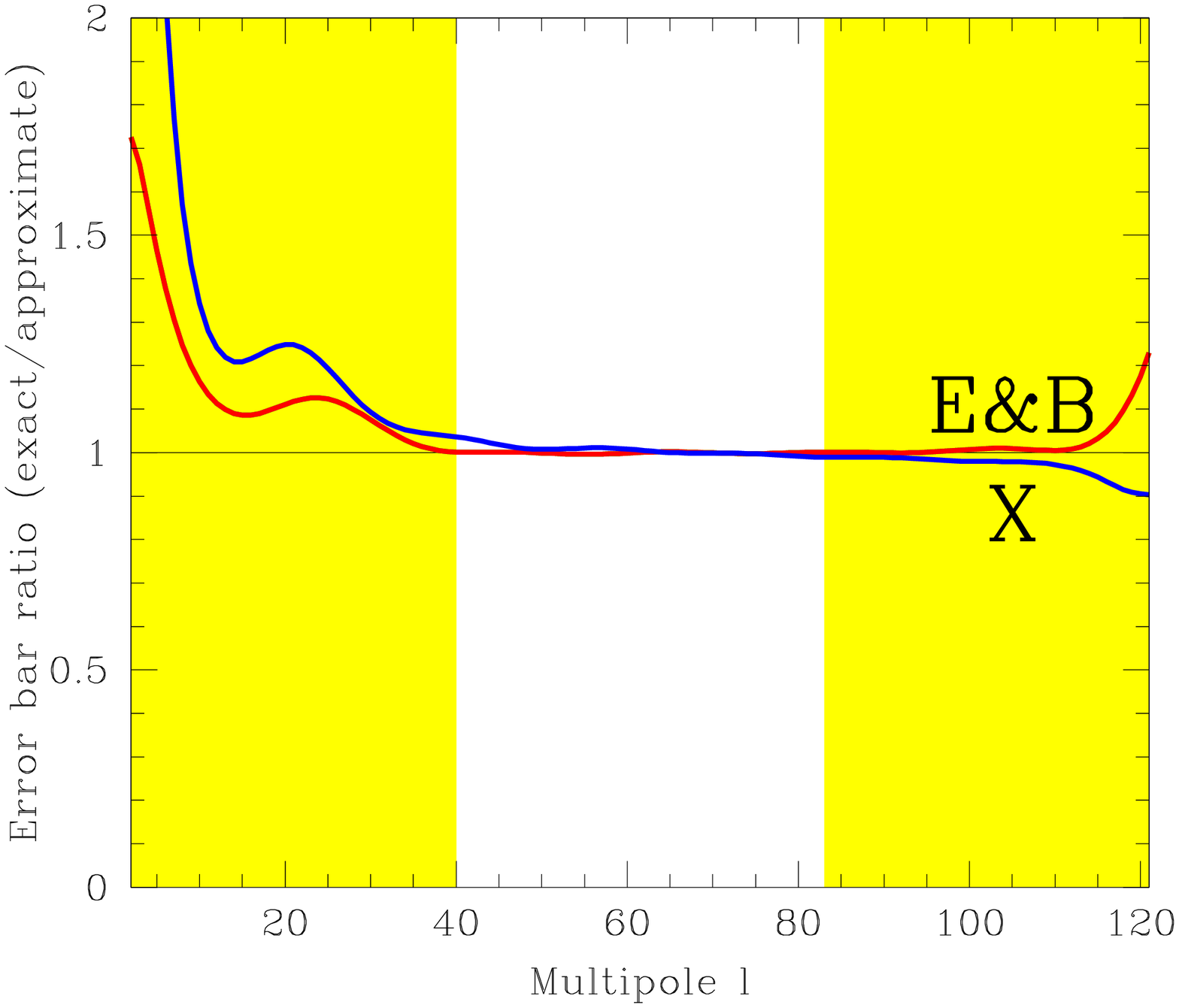}}
\postskip
\caption{\label{ErrorFig}\footnotesize%
The curves show the ratio of the actual error bars to
those computed with the approximation of \eq{PowerCovEq}.
The approximation is seen to be excellent 
for multipoles more than a couple of bandwidths $\Delta\l$ 
away from the two endpoints of the $\l$-range computed 
(outside shaded regions).
}
\end{figure}

The approximation that the number of uncorrelated modes
scales as $\fsky$ is both natural and well-motivated \cite{ScottCosmicVar}.
How accurate is it in practice?
Our calculations enable us to address this issue quantitatively.
\Fig{ErrorFig} shows the ratio of the approximate error bars
from \eq{PowerCovEq} to the exact error bars from \eq{qCovEq}
for the four power spectra (the ratios of the square roots of
the corresponding elements on the covariance matrix diagonals). 
To ensure a fair comparison, 
we used the uncorrelated method given by
\eq{QdefEq} with $\B=\F^{-1/2}$ --- the $\B=\F^{-1}$ and maximum-likelihood
methods give larger anticorrelated error bars, whereas the
$\B=\I$ method gives smaller correlated ones.\footnote{
The reader interested in implementing any of these methods in practice should note that
care needs to be taken when the bands used are much narrower than $\Delta\l$, since
this makes $\F$ for all practical purposes singular, with many eigenvalues 
so close to zero that rounding errors make them slightly negative.
Our B2001 example with bandwidth 1 (we are measuring for each $\l$ separately
and $\Delta\l\sim 30$) is a case in point.
For the $\B=\F^{-1/2}$ case, the window functions are simply proportional to the rows of
$\F^{1/2}$, so they are readily computed by setting the offending eigenvalues to zero.
This determines the normalization constants as
$\norm_i = 1/\sum_{j=1}^\Nb(\F^{1/2})_{ij}$, since each window function must sum to unity.
The error bar $\Delta\q_i$ is then equal to $\norm_i$.
In short, all plots in this paper remain well-defined even when $\F$ is singular.
Evaluating $\q$ in practice, however, is not possible when 
$\F$ is singular, since it involves calculating $\F^{-1/2}$ numerically.
If $\B$ is chosen to be a regularized version of  $\F^{-1/2}$ in 
\eq{QdefEq}, the decorrelation method fails in the sense that the 
band power estimates $q_i$ will exhibit a slight residual correlation.
In conclusion, the power spectrum should not be too oversampled
for actual data analysis. The same obviously applies to the 
$\B=\F^{-1}$ method.
}
It is seen that the approximation of \eq{PowerCovEq} is generally quite 
accurate when $\l\gg\Delta\l$ and $\l\ll\lmax-\Delta\l$, 
\ie, for multipoles well away from the two endpoints
of the range computed.
This means that forecasts made using the approximation of 
\eq{PowerCovEq} should be viewed 
as quite accurate except on scales comparable to the survey size.

For the case where all six power spectra are measured jointly, 
\eq{PowerCovEq} is generalized so that the matrix in parenthesis gets
replaced by the following one (suppressing the subscript $\l$ for brevity):
\beq{PowerCovEq2}
\left(\begin{tabular}{cccccc}
$T^2$	&$X^2$	&$Y^2$	&$TX$		&$TY$		&$XY$\\[2pt]
$X^2$	&$E^2$	&$Z^2$	&$EX$		&$XZ$		&$EZ$\\[2pt]
$Y^2$	&$Z^2$	&$B^2$	&$YZ$		&$BY$		&$BZ$\\[2pt]
$TX$	&$EX$	&$YZ$	&${TE+X^2\over2}$	&${TZ+XY\over 2}$	&${EY+XZ\over 2}$\\[2pt]
$TY$	&$XZ$	&$BY$	&${TZ+XY\over 2}$	&${TB+Y^2\over2}$	&${BX+YZ\over 2}$\\[2pt]
$XY$	&$EZ$	&$BZ$	&$\>{EY+XZ\over 2}$	&$\>{BX+YZ\over 2}$	&$\>{EB+Z^2\over2}$
\end{tabular}\right)
\eeq
This can either be derived directly from a quadratic estimator as in \cite{ZSS97}
or by computing the inverse Fisher information matrix as in \cite{foregpars}.

\section{Discussion}
\label{DiscussionSec}

We have presented a method for measuring CMB polarization power spectra 
given incomplete sky coverage and tested it with a number of 
simulated examples.

\subsection{What have we learned?}

The issue of measuring the {\bf T power spectrum} has 
been extensively treated in prior literature. 
An added challenge when measuring the {\bf E and B power spectra}
is leakage between the two caused by incomplete sky coverage. 
We quantified this leakage for the first time, and
found that it is rather
insensitive to experimental noise levels and depends mainly on 
geometry. Specifically, we found the leakage to depend mainly 
on the ratio $\l/\Delta\l$, where $\Delta\l$ is the characteristic
window function width and scales roughly as the inverse 
size of the sky patch in its narrowest direction.
We introduced a disentanglement method which 
reduced the leakage to $5\%-10\%$ for $\l\simgt\Delta\l$,
and described how it could be pushed to zero if necessary, at
the cost of larger error bars.

We found that when measuring the {\bf X power spectrum},
the basic quadratic estimator produces a surprisingly 
complicated answer, involving not only temperature-polarization
cross terms, but also, \eg, autocorrelations of the 
temperature map with itself. This is unfortunate,
since it can make the measured power spectrum more sensitive to
systematic errors. Especially if the temperature and polarization
maps are made by two different experiments, systematics should
be uncorrelated between the two and therefore not contribute to
the cross term average.
The maximum-likelihood method exhibits the same problem.
This can affect $E$- and $B$-polarization estimation as well,
by giving non-zero weight to the unpolarized map.
The problem is eliminated by simply using vanishing 
cross-power in the fiducial model.
We find that this is desirable in practice, since the resulting 
information loss causes error bars to increase only by
negligle amounts, at the percent level.

Finally, we found that on scales
substantially smaller than the sky patch,
the error bars for the $F^{-1/2}$-method
were accurately fit by the approximation of \eq{PowerCovEq},
where variance scales inversely with sky coverage.

\subsection{Relation to other methods}

The quadratic estimator (QE) method is closely related to 
the maximum-likelihood (ML) method: the latter is simply
the quadratic estimator method with $\B=\F^{-1}$ in \eq{QdefEq},
iterated so that the fiducial (``prior'') power spectrum 
equals the measured one \cite{BJK}.
The ML method has the advantage of not requiring any prior to be assumed.
The QE method has the advantage of being faster (no iteration) and 
simpler to interpret --- since it is quadratic rather than highly non-linear,
the statistical properties the measured band power vector $\q$ can
be computed analytically. This allows the likelihood function to be computed
directly from $\q$ (as opposed to $\x$), in terms of generalized 
$\chi^2$-distributions \cite{Wandeldt}.

Both methods are unbiased, but they may differ as regards error bars.
The QE method can produce inaccurate error bars if the prior is
inconsistent with the actual measurement.
The ML method Fisher matrix can produce inaccurate error bar estimates
if the measured power spectra have substantial scatter due to noise or 
sample variance, in which case they are unlikely to describe the 
smoother true spectra.
A good compromise is therefore to iterate the QE method once and choose 
the second prior to be a rather smooth model consistent with
the original measurement. In addition, as mentioned above, we found that it
is useful to set the $X$ power spectrum to zero in the prior.

The difference between the QE and ML methods is often small in 
practice, which can be understood as follows. 
Since the QE method can be shown to be lossless if 
the prior equals the truth, thereby minimizing the error bars,
small departures from the true prior merely 
destroy information to second order. This is also
why adopting a prior without $X$ power inflated the
error bars so little.

The analysis of weak gravitational lensing data is rather analogous
to that of CMB polarization, since the projected shear field can
be decomposed into components corresponding to $E$ and $B$.
Recent lensing work cast in this language has included both 
theoretical predictions for various effects 
\cite{Benabed99,Benabed00} and data analysis issues
\cite{HuWhite00,Crittenden00}.
In particular, no $E/B$ leakage was detected at the 10\% level
when the ML method was applied to 
simulated examples \cite{HuWhite00}, which can be understood
from our present results 
since the band power bins used were substantially broader than 
$\Delta\l$\cite{HuWhite00} and mainly on scales $\l\gg\Delta\l$.

\subsection{Outlook}

Our basic results are good news: although polarized power spectrum
estimation adds several complications to the non-polarized case, 
they can all be dealt with using the techniques we have described.
However, much theoretical work remains to be done. Here are a couple of examples
of areas deserving further study:
\begin{itemize}
\item 
The effects of pixel shape and size in the mapmaking process needs to be quantified,
and is likely to be more important for the polarized case.
\item
Although the methods we have discussed apply equally well 
to measuring the power spectra of non-Gaussian signals 
(the only change is that the error bars will no longer be
given by \eq{qCovEq}), non-Gaussian signals
contain more information than is contained in their 
power spectra. 
Polarized non-Gaussian fluctuations are expected from 
microwave foregrounds \cite{Kogut00,Tucci00,Baccigalupi00},
secondary anisotropies \cite{Hu00}, 
topological defects \cite{Benabed99} and
CMB lensing \cite{Benabed00}, and only a small
number of papers have so far addressed the issue of how to 
best measure such non-Gaussian signatures 
in practice \cite{Popa98,Dolgov99,Kotok00}.
\end{itemize}
First and foremost, however, we need a detection of CMB polarization!

\bigskip
The authors wish to thank Wayne Hu, Brian Keating, Ue-Li Pen and Matias Zaldarriaga 
for helpful comments, and the B2001, PIQUE and POLAR teams for encouraging us
to write up these calculations.
Support for this work was provided by
NASA grant NAG5-9194,
NSF grant AST00-71213 and 
the University of Pennsylvania Research Foundation.

\clearpage
\appendix

\section{Computing the polarization covariance matrix}

This Appendix is intended for the reader who wishes to 
write software to explicitly compute the polarization 
covariance matrix.
The complete formalism for describing CMB polarization was presented in 
\cite{Kamion97,Zalda97}, and extended with a number of useful 
explicit formulas in \cite{Zalda98}. However, a number
of practical details are not covered in the literature, \eg,
the rotation angles and degenerate cases, so we describe
all steps explicitly for completeness. 

Let the 3-dimensional vector $\x_i$ denote the three
measurable quantities for the $\ith$ pixel:
\beq{xiDefEq}
\x_i\equiv\left(\begin{tabular}{c}
$T_i$\\
$Q_i$\\
$U_i$
\end{tabular}\right).
\eeq
The $3\times 3$ covariance matrix between two such vectors
at different points can be written 
\beq{BasicCovarEq}
\expec{\x_i\x_j^t}
= \R(\alpha_{ij})\M(\rh_i\cdot\rh_j)\R(\alpha_{ji})^t,
\eeq
Here $\M$ is the covariance using a $(Q,U)$-convention where the 
reference direction is the great circle connecting the two points,
and the rotation matrices given by
\beq{RdefEq}
\R(\alpha)\equiv
\left(\begin{tabular}{ccc}
$1$	&$0$			&$0$\\[2pt]
$0$	&$\cos 2\alpha$		&$\sin 2\alpha$\\[2pt]
$0$	&$-\sin 2\alpha$	&$\cos 2\alpha$\\[2pt]
\end{tabular}\right)
\eeq
accomplish a rotation
into a global reference frame where the reference directions are meridians.
The full $(3n)\times(3n)$ map covariance matrix is readily assembled out of the 
$3\times 3$ blocks of \eq{BasicCovarEq} by looping over all pixel pairs.
A powerful probe for bugs is making sure that this large matrix is positive definite.

\subsection{Computing the rotation angles}

As we will see, computing the magnitudes of the rotation angles
$\alpha_{ij}$
is straightforward, whereas getting the correct sign is somewhat subtle.

The great circle connecting the two pixels has the unit normal vector
\beq{rijEq}
\rh_{ij}\equiv{\rh_i\times\rh_j\over|\rh_i\times\rh_j|}.
\eeq
Similarly,  
the meridian passing though pixel $i$ (its reference circle for our global
$(Q,U)$-convention) has the unit normal vector
\beq{ristarEq}
\rh_i^*\equiv{\zh\times\rh_i\over|\zh\times\rh_i|},
\eeq
where $\zh=(0,0,1)$ is the unit vector in the $z$-direction.
The magnitude of $\alpha_{ij}$, the rotation angle for pixel $i$,
is simply the angle between these two great circles, so
$\cos\alpha_{ij} = \rh_{ij}\cdot\rh_i^*$.
The sign of $\alpha_{ij}$ is defined so that 
a positive angle corresponds to 
clockwise rotation at the pixel (at $\rh_i$). 
We therefore compute the cross product of the
two circle normals, 
which has the property that
\beq{sinalphaEq}
\rh_{ij}\times\rh_i^* = \rh_i\sin\alpha_{ij}.
\eeq
(Since both $\rh_{ij}$ and $\rh_i^*$ are by construction
perpendicular to $\rh_i$, their cross product will be either
aligned or anti-aligned with $\rh_i$.)
Dotting \eq{sinalphaEq} with $\rh_i$ and performing some
vector algebra gives 
\beqa{sinalphaEq2}
\sin\alpha_{ij}&=&(\rh_{ij}\times\rh_i^*)\cdot\rh_i
\propto[(\rh_i\times\rh_j)\times (\rh_i\times\zh)]\cdot\rh_i\nonumber\\
&=&[(\rh_i\times\rh_j)\cdot\zh]\>\rh_i\cdot\rh_i
\propto\rh_{ij}\cdot\zh,
\eeqa
where the omitted proportionality constants are positive.
$\alpha_{ij}$ therefore has the same sign as the $z$-coordinate of 
$\rh_{ij}$, and is given by 
\beq{alphaEq}
\alpha_{ij} = \cases{
+\cos^{-1}\left(\rh_{ij}\cdot\rh_i^*\right)	&if $\rh_{ij}\cdot\zh>0$,\crr
-\cos^{-1}\left(\rh_{ij}\cdot\rh_i^*\right)	&if $\rh_{ij}\cdot\zh<0$.
}
\eeq




For generic pairs of directions, \eq{alphaEq} gives the two rotation angles 
$\alpha_{ij}$ and $\alpha_{ji}$ needed for \eq{BasicCovarEq}.
However, it breaks down for the three special cases 
$\rh_i\times\rh_j=\zero$,
$\rh_i\times\zh=\zero$ and
$\rh_j\times\zh=\zero$.
If $\rh_i\times\rh_j=\zero$, the two pixels are either identical
or on diametrically opposite sides of the sky.
Hence any great circle through $\rh_i$ will go through $\rh_2$ as well.
We can choose this circle
to be the meridian, so no rotation is needed, \ie, 
$\alpha_{ij}=\alpha_{ji}=0$ for this case.
Indeed, $\M$ comes out diagonal for this case by symmetry, with 
$\expec{Q_i U_j}=0$ and 
$\expec{Q_i Q_j}=\expec{U_i U_j}$, so rotations have no effect.

If $\rh_i\times\zh=\zero$, then the pixel is at the North or South pole,
making our the global $(Q,U)$-convention undefined.
The simplest remedy to this problem is to move the pixel away from
the pole by a tiny amount much smaller than the beam width of the 
experiment.

The remainder of this section discusses some symmetry issues 
that the pragmatic reader may wish to skip.
\Eq{alphaEq} guarantees a symmetric covariance matrix since
swapping $i$ and $j$ is equivalent to transposing the result.
Moreover, the two rotations 
$\R(\alpha_{ij})$ and $\R(\alpha_{ji})$ 
are nearly equal when the 
two pixels are much closer to each other than to the poles
(the two rotations would be identical for a flat sky), 
which can be seen as follows.
$\rh_{ji}=-\rh_{ij}$ from the antisymmetry of the cross product.
Since $\rh_i\approx\rh_j$ implies that $\rh_i^*\approx\rh_j^*$,
and $\cos^{-1}(-x)=\pi-\cos^{-1}(x)$, 
\eq{alphaEq} therefore gives 
$\alpha_{ji}\approx -(\pi-\alpha_{ij}) = \alpha_{ij} - \pi.$
The extra rotation by $\pi$ has no effect, since the rotation matrix
of \eq{RdefEq} depends on twice the angle.

There is a more subtle symmetry as well. 
Flipping the overall sign in \eq{alphaEq}
will give the same covariance matrix if we redefine $U$ with the 
opposite sign convention.
The sign convention of \eq{alphaEq} corresponds to the standard 
definition of $U$.

\subsection{Computing the matrix $\M$}

The $\M$-matrix depends only on the angular separation between 
the two pixels. It is given by
\beq{MdefEq}
\M(\rh_i\cdot\rh_j)\equiv
\left(\begin{tabular}{ccc}
$\expec{T_i T_j}$	&$\expec{T_i Q_j}$	&$\expec{T_i U_j}$\\[2pt]
$\expec{T_i Q_j}$	&$\expec{Q_i Q_j}$	&$\expec{Q_i U_j}$\\[2pt]
$\expec{T_i U_j}$	&$\expec{U_i Q_j}$	&$\expec{U_i U_j}$\\[2pt]
\end{tabular}\right),
\eeq
\beqa{CorrEq}
\expec{T_i T_j}&\equiv&
  \sum_\l \left({2\l+1\over 4\pi}\right)P_\l(z) C_\l^T,\\
\expec{T_i Q_j}&\equiv&
  -\sum_\l \left({2\l+1\over 4\pi}\right)
  F^{10}_\l(z) C_\l^{TE},\\
\expec{T_i U_j}&\equiv&
  -\sum_\l \left({2\l+1\over 4\pi}\right)
  F^{10}_\l(z) C_\l^{BT},\\
\expec{Q_i Q_j}&\equiv&
  \sum_\l \left({2\l+1\over 4\pi}\right)
  \left[F^{12}_\l(z) C_\l^E - F^{22}_\l(z) C_\l^B\right],\\
\expec{U_i U_j}&\equiv&
  \sum_\l \left({2\l+1\over 4\pi}\right)
  \left[F^{12}_\l(z) C_\l^B - F^{22}_\l(z) C_\l^E\right],\\
\expec{Q_i U_j}&\equiv&
  \sum_\l \left({2\l+1\over 4\pi}\right)
  \left[F^{12}_\l(z) + F^{22}_\l(z)\right] C_\l^{EB},\label{CorrEq6}
\eeqa
where $\z=\rh_i\cdot\rh_j$ is the cosine of the 
angle between the two pixels 
under consideration.
The equations for $\expec{T_i Q_j}$, $\expec{Q_i Q_j}$ and $\expec{U_i U_j}$ are from 
\cite{Zalda98} (beware a minus sign typo in the first one) and those for 
$\expec{T_i U_j}$ and $\expec{Q_i U_j}$ are from \cite{Zalda01}.
$P_\l$ denotes a Legendre polynomial, and 
the functions $F^{10}$, $F^{12}$ and $F^{22}$ are \cite{Zalda98}
\beqa{FdefEq}
F^{10}(z)&=&
2{
{\l z\over(1-z^2)} P_{\l-1}(z)
-\left({\l\over 1-z^2} + {\l(\l-1)\over 2}\right) P_\l(z)
\over
[(\l-1)\l(\l+1)(\l+2)]^{1/2}
}\\\crr
\label{FdefEq2}
F^{12}(z)&=&
2{
{(\l+2)z\over(1-z^2)} P_{\l-1}^2(z)
-\left({\l-4\over 1-z^2} + {\l(\l-1)\over 2}\right) P_\l^2(z)
\over
(\l-1)\l(\l+1)(\l+2)
}\\\crr
\label{FdefEq3}
F^{22}(z)&=&
4{
(\l+2) P_{\l-1}^2(z)
-(\l-1)z P_\l^2(z)
\over
(\l-1)\l(\l+1)(\l+2)(1-z^2)
}
\eeqa
Here $P_\l$ and $P_\l^2$ denote Legendre polynomials $P_\l^m$ for the
cases $m=0$ and $m=2$, respectively, which are 
conveniently computed using the recursion relations
\cite{GradshteynRyzhik}
\beqa{RecursionEq}
P_\l(z) &=&\cases{
1							&for $\l=0$,\crr
z							&for $\l=1$,\crr
{(2\l-1) z P_{\l-1}(z) - (\l-1)P_{\l-2}(z)\over\l}	&for $\l\ge 2$,
}\\
P_\l^2(z) &=&\cases{
3(1-z^2)						&for $\l=2$,\crr
5z P_2^2							&for $\l=3$,\crr
{(2\l-1) z P_{\l-1}^2(z) - (\l+1)P_{\l-2}^2(z)\over\l-2}&for $\l\ge 4$.
}
\eeqa
The division by $(1-z^2)$ unfortunately causes expressions 
\eqn{FdefEq}$-$(\ref{FdefEq3})
to blow up numerically when
$z=\pm 1$, \ie, for correlations at zero separation or between
diametrically opposite pixels in the sky \cite{Ng}.
Taking the appropriate limits for these cases gives 
\beqa{FlimitEq}
F_\l^{10}(z)&=&0\quad\quad\quad\quad\>\>\,\hbox{if $|z|=1$},\\\crr
\label{FlimitEq2}
F_\l^{12}(z)&=& \cases{
{1\over 2}		&if $z=1$,\crr
{1\over 2}(-1)^\l	&if $z=-1$,
}\\\crr
\label{FlimitEq3}
F_\l^{22}(z)&=& \cases{
-{1\over 2}		&if $z=1$,\crr
{1\over 2}(-1)^\l	&if $z=-1$.
}
\eeqa

\subsection{Including variable angular resolution}

If the experimental beam is rotationally symmetric,
its effect is straightforward to include even if the beam 
size and radial profile varies between pixels. This complication 
is particularly important for the case of cross-polarization,
where it may be desirable to correlate polarization maps 
from one experiment with a temperature map from another experiment
that happens to have different angular resolution.

Let $B_{i\l}$ denote the coefficients obtained from expanding the 
beam profile corresponding to the measurement $x_i$ in Legendre polynomials.
For a Gaussian beam, these coefficients are accurately described by
the well-known approximation
\beq{GaussBeamEq}
B_{i\l} \approx e^{-{1\over 2}\theta_i^2\l(\l+1)},
\eeq
where the rms beam width $\theta_i$ is related to the
full-width-half-maximum by $\theta_i=$FWHM$_i/\sqrt{8\ln 2}$.
To compute the correlation $\expec{x_i x_j}$ in 
\eq{BasicCovarEq},
the six power spectra $C_\l^P$ are simply replaced
by $C_\l^P B_{i\l} B_{j\l}$ in 
equations\eqn{CorrEq} through\eqn{CorrEq6}.



\end{document}